\documentclass[10pt,fleqn]{article}
\pdfoutput=1
\usepackage{psfrag}
\usepackage{amsmath}
\usepackage[dvips]{epsfig}
\usepackage{setspace}
\usepackage{amsmath,amssymb,bm}
\usepackage{graphics}
\usepackage[pdftex,colorlinks,bookmarks,pdfpagemode=UseOutlines,linkcolor=black,
urlcolor=black,citecolor=black,pageanchor=false]{hyperref}
\usepackage[margin=1in]{geometry}
\usepackage[T1]{fontenc}
\usepackage{fix-cm}
\usepackage[authoryear,round]{natbib}
\usepackage[pagewise,mathlines]{lineno}
\usepackage{amsthm}
\usepackage{longtable}
\usepackage{xfrac}
\usepackage{booktabs}
\usepackage{multirow}

\graphicspath{{figures/}{Figures/}}

\usepackage{placeins}
\let\Oldsection\section
\renewcommand{\section}{\FloatBarrier\Oldsection}

\let\Oldsubsection\subsection
\renewcommand{\subsection}{\FloatBarrier\Oldsubsection}

\newtheorem{theo}{Theorem}[section]
\newtheorem{lemma}{Lemma}[section]
\newtheorem{df}{Definition}[section]
\newtheorem{cor}{Corollary}[section]
\newtheorem{assump}{Assumption}[section]
\newtheorem{assert}{Assertion}[section]
\newtheorem{remark}{Remark}[section]
\newcommand{\bl}{\begin{lemma}}
\newcommand{\el}{\end{lemma}}
\newcommand{\be}{\begin{equation}}
\newcommand{\ee}{\end{equation}}
\newcommand{\beqn}{\begin{eqnarray}}
\newcommand{\eeqn}{\end{eqnarray}}
\newcommand{\bt}{\begin{theo}}
\newcommand{\et}{\end{theo}}
\newcommand{\bd}{\begin{df}}
\newcommand{\ed}{\end{df}}
\newcommand{\ba}{\begin{assump}}
\newcommand{\ea}{\end{assump}}
\newcommand{\bass}{\begin{assert}}
\newcommand{\eass}{\end{assert}}
\newcommand{\brem}{\begin{remark}}
\newcommand{\erem}{\end{remark}}
\newcommand{\bc}{\begin{cor}}
\newcommand{\ec}{\end{cor}}

\newcommand{\BB}{{\cal B}}
\newcommand{\DD}{{\cal D}}

\newcommand{\Lcal}{{\cal L}}

\newcommand{\Ncal}{{\cal N}}

\newcommand{\pt}{\tilde{p}}

\newcommand{\St}{\tilde{S}}
\newcommand{\So}{\overline{S}}

\newcommand{\Ut}{\tilde{U}}
\newcommand{\Vt}{\tilde{V}}
\newcommand{\OBB}{\overline{\BB}}

\newcommand{\etab}{\bm{\eta}}

\newcommand{\MLnorm}{\textnormal{ML}}

\numberwithin{equation}{section}

\long\def\comment#1{}

\title{Macroscopic Modeling, Calibration, and Simulation of Managed Lane-Freeway Networks, Part II: Network-scale Calibration and Case Studies
\author{Matthew A. Wright, Roberto Horowitz, Alex A. Kurzhanskiy}
\date{}
}

\begin{document}
\maketitle

\begin{abstract}
In Part I of this paper series, several macroscopic traffic model elements for mathematically describing freeway networks equipped with managed lane facilities were proposed.
These modeling techniques seek to capture at the macroscopic the complex phenomena that occur on managed lane-freeway networks, where two parallel traffic flows interact with each other both in the physical sense (how and where cars flow between the two lane groups) and the physiological sense (how driving behaviors are changed by being adjacent to a quantitatively and qualitatively different traffic flow).

The local descriptions we developed in Part I are not the only modeling complexity introduced in managed lane-freeway networks.
The complex topologies mean that network-scale modeling of a freeway corridor is increased in complexity as well.
The already-difficult model calibration problem for a dynamic model of a freeway becomes more complex when the freeway becomes, in effect, two interrelating flow streams.
In the present paper, we present an iterative-learning-based approach to calibrating our model's physical and driver-behavioral parameters.
We consider the common situation where a complex traffic model needs to be calibrated to recreate real-world baseline traffic behavior, such that counterfactuals can be generated by training purposes.
Our method is used to identify traditional freeway parameters as well as the proposed parameters that describe managed lane-freeway-network-specific behaviors.
We validate our model and calibration methodology with case studies of simulations of two managed lane-equipped California freeways.
\end{abstract}

{\bf Keywords}: macroscopic first order traffic model, first order node model,
multi-commodity traffic, managed lanes, HOV lanes, dynamic traffic assignment,
dynamic network loading, inertia effect, friction effect, smoothing effect

\section{Introduction}\label{hov2:sec_intro}
Managed lanes \citep{obenberger_managed_2004}, such as high-occupancy-vehicle (HOV) lanes or tolled lanes, have become a popular policy tool for transportation authorities, who seek to capture both demand management outcomes by incentivizing behaviors like carpooling \citep{chang_review_2008} and to gain additional real-time control ability \citep{kurzhanskiy_traffic_2015}.

In Part I of this paper series \citep{wright_macroscopic_2019}, we proposed several macroscopic modeling elements for the purpose of mathematically describing a freeway equipped with a managed lane facility (a ``managed lane-freeway network'').
Part I discusses the road-topological considerations of modeling two parallel traffic flows, as well as mathematical descriptions for several of the complex emergent (and sometimes counteracting) phenomena that have been observed on managed lane-freeway networks, like the so-called friction \citep{liu_analysis_2011,jang_traffic_2012,fitzpatrick_operating_2017} and smoothing \citep{menendez_effects_2007, cassidy_smoothing_2010,jang_dual_2012} effects.

The present paper discusses the actual implementation of these modeling constructions.
Transportation authorities considering a significant infrastructure investment like building new roads or installing managed lane facilities will make use of simulation tools during the pre-construction planning phase \citep{how_caltrans_builds_projects}.
Models let transportation planners understand the characteristics of the road network, and predict outcomes of proposed modifications before their undertaking.

The use of models for the predictive study of counterfactuals is especially relevant for managed lane deployments.
A major benefit of modern managed lanes to transportation authorities is how they can enable real-time, reactive control \citep{obenberger_managed_2004,kurzhanskiy_traffic_2015}.
Managed lanes today may be instrumented with operational capabilities for real-time traffic control, such as dynamic, condition-responsive toll rates in tolled express lanes \citep{lou_optimal_2011}.
Real-time control decisions can be studied via a well-tuned predictive model.
As the traffic situation changes, multiple potential operational strategies can be evaluated to decide on the best course of action.

Before a traffic model can be used for forecasting or generating counterfactuals, it must be tuned to represent a base-case scenario, usually the existing traffic patterns.
This process of model tuning is generally referred to as \emph{calibration}.
Calibration of traffic network models is known to be a difficult task \citep[Chapter 16]{treiber_traffic_2013}:
vehicle traffic dynamics exhibit many complex and interacting nonlinear behaviors, and a model must realistically capture these behaviors to be useful for analysis or planning purposes.
For a freeway-corridor-scale network, the data available for model calibration may consist of vehicle counts and mean speeds from vehicle detectors, origin-destination survey data, and/or descriptive information about the regular spatiotemporal extent of regular.
A calibration process involves tuning a simulation model's parameters such that it accurately reproduces typical traffic behaviors \citep{treiber_traffic_2013}.
The nonlinear and chaotic nature of traffic means that calibrated parameter values typically cannot be found explicitly, and must be sought after via an iterative tuning-assessment process.

In this paper, we present a calibration methodology for our recently-developed macroscopic managed lane-freeway network modeling constructions \citep{wright_macroscopic_2019}, that integrates them into the broader macroscopic traffic model calibration loop.
Much of our focus is on identifying the split ratios, the portions of vehicles that take each of several available turns between the managed lane(s) and the general-purpose (GP) lane(s).
These values are of particular importance when the behavior of the managed lane is of interest, and have major effects on the resulting analysis of managed lane usage and traffic behaviors, but are not directly measurable with traditional traffic sensors.
Instead, they must be found as part of the iterative calibration process such that the macroscopic traffic patterns actually observed on the managed lane's instrumentation are reproduced in the model.
A managed lane-freeway network calibration procedure, in other words, is a \emph{superset} of the calibration procedure for a typical freeway corridor model.

The remainder of this paper is organized as follows.
Section~\ref{hov2:sec_full_model} outlines the macroscopic traffic model in question, which augments the traditional kinematic-wave macroscopic traffic model with the managed-lane-specific constructions developed in \citet{wright_macroscopic_2019}.
Calibration procedures for the managed-lane-specific components, and the integration of them into a full calibration loop, is discussed in Section~\ref{hov2:sec_calibration}.
Two case studies of macroscopic modeling of managed-lane-equipped freeways in California, one with a full-access managed lane and one with a separated managed lane with gated access, are presented in Section~\ref{hov2:sec_simulation}, followed by concluding points in Section~\ref{hov2:sec_conclusion}

\newcommand{\obeta}{\overline{\beta}}
\newcommand{\tbeta}{\tilde{\beta}}

\section{A Managed Lane-Freeway Network Simulation Model}\label{hov2:sec_full_model}
In this Section, we present a macroscopic simulation algorithm for managed lane-freeway networks.
This can be considered a fleshing-out of the simplified first-order macroscopic simulation method briefly outlined as context in \citet{wright_macroscopic_2019}, with extensions made by incorporating the additional items we described in the same paper.

\subsection{Background}
``Macroscopic'' traffic models are those that model traffic flows via the continuum fluid approximation (originally proposed by \citet{lighthill_kinematic_1955_pt1, lighthill_kinematic_1955_pt2,richards_shock_1956}) as opposed to ``microscopic'' models that model individual vehicles (\citet{treiber_traffic_2013} present a relatively recent broad survey of various types of traffic models).
While macroscopic models necessarily have a lower resolution than microscopic models, they make up for it in areas like shorter runtimes (making them useful for real-time operational analysis).

The present paper considers a macroscopic model for managed lane-freeway networks, based on the family of  ``first-order,'' ``kinematic wave''-type models descending from the Cell Transmission Model \citep{daganzo_cell_1994,daganzo_cell_1995}.

\subsection{Definitions}
\begin{itemize}
  \item We have a network consisting of set of directed links $\Lcal$, representing segments of road, and a set of nodes $\Ncal$, which join the links.
  \begin{itemize}
    \item A node always has at least one incoming link and one outgoing link.
    \item A link may have an upstream node, a downstream node, or both.
  \end{itemize}
  \item We have $C$ different vehicle classes traveling in the network, with classes indexed by $c \in \{1,\dots,C\}$.
  \item Let $t \in \{0,\dots,T\}$ denote the simulation timestep.
  \item In addition, while we have not covered them here, the modeler may optionally choose to define \emph{control inputs} to the simulated system that modify system parameters or the system state.
  Such control inputs may represent operational traffic control schemes such as ramp metering, changeable message signs, a variable managed-lane policy, etc.
  In the context of this paper, we suggest including the parameterization of the friction effect and the class-switching construction of the separated-access managed lane model to be discussed in section \ref{hov2:subsec_beta_sep} as control actions.
\end{itemize}

\subsubsection{Link model definitions}
Traditionally, the mathematical equations that govern the traffic flow in the links are called the ``link model.''
\begin{itemize}
	\item For each link $l \in \Lcal$, let there be a time-varying $C$-dimensional state vector $\vec{\rho}_l(t)$, which denotes the density of the link of each of the $C$ vehicle classes at time $t$.
  \begin{itemize}
    \item Each element of this vector, $\rho_l^c(t)$, updates between timestep $t$ and timestep $t+1$ according to \eqref{hov2:eq:update}.
    \item Also define $\rho_{l,0}^c$ for all $l,c$, the initial condition of the system.
  \end{itemize}
  \item We define three types of links:
    \begin{itemize}
      \item Ordinary links are those links that have both beginning and ending nodes.
      \item Origin links are those links that have only an ending node.
        These links represent the roads that vehicles use to enter the network.
      \item Destination links are those links that have only a beginning node.
        These links represent the roads that vehicles use to exit the network.
    \end{itemize}
  \item For each link $l \in \Lcal$, define the actual parametric ``link model'' that computes the per-class demands $S_l^c(t) \leq \rho_l^c(t)$ (the amount of vehicles that want to exit the link at time $t$) and link supply $R_l(t)$ (the amount of vehicles that link $l$ can accept at time $t$) as functions of $t$ and $\vec{\rho}_l(t)$.
  The particular link model equations is a modeling choice, and many authors have proposed different versions.
  Appendix \ref{hov:app_linkmodel} describes a particular example link model that will be used in the example simulations in Section \ref{hov2:sec_simulation}.
  \item For each origin link $l \in \Lcal$, define a $C$-dimensional time-varying vector, $\vec{d}_l(t)$, where the $c$-th element $d_l^c(t)$ denotes the exogenous demand of class $c$ into the network at link $l$.
\end{itemize}

\subsubsection{Node model definitions}
Analogous to the link model, the equations that govern the traffic flows through nodes (i.e., between links) are called the ``node model.''
\begin{itemize}
  \item For each node $\nu \in \Ncal$, let $i \in \{1,\dots,M_\nu\}$ denote the incoming links and $j \in \{1,\dots,N_\nu\}$ denote the outgoing links.
  \item For each node, define $\{ \beta_{ij}^c(t): \; \sum_j \beta_{ij}^c(t) = 1 \; \forall i,c\}$ the time-varying split ratios for each triplet $\{i,j,c\}$.
Each split ratio may be fully defined, partially defined, or fully undefined, with the undefined split ratios typically being the ones that specify the crossflows between the managed lane(s) and the GP lane(s).
\item For each node with both a managed lane link and a GP link exiting it, define a (state-dependent) split ratio solver for the managed lane-eligible vehicles.
We suppose that the decision of eligible vehicles of whether or not to change between the two lane types is dependent on the local traffic conditions.
Appropriate methods for determining the desired lane type as a function of the current state may include logit-style discrete choice methods \citep{farhi_logit_2013} or dynamic-system-based models such as the one proposed in \citet{wright_dynamic-system-based_2018} and reviewed in Appendix \ref{hov:app_splitratiosolver}.
  \item For each node, define a ``node model'' that, at each time $t$, takes its incoming links' demands $S_i^c(t)$ and split ratios $\beta_{ij}^c(t)$, its outgoing links' supplies $R_j(t)$, and other nodal parameters, and computes the flows $f_{ij}^c(t)$.
  As with the link model, the particular node model equations is a modeling choice, and much literature exists defining and analyzing different node models.
  In this paper, we refer to a specific node model with a relaxed first-in-first-out (FIFO) construction that has additional parameters $\etab_{j'j}^i(t)$ (mutual restriction intervals) and $p_i(t)$ (the incoming links' priorities).
  Note also that a node model with the relaxed FIFO construction can also be used to produce the smoothing effect of managed lanes, as discussed in depth in Part I \citep{wright_macroscopic_2019}.
\end{itemize}

\subsubsection{State update equation definitions}
These equations define the time evolution of the states from time $t$ to time $t+1$.
\begin{itemize}
  \item All links $l \in \Lcal$ update their states according to the equation
  \begin{linenomath}
  \begin{equation}
    \rho_l^c(t+1) = \rho_l^c(t) + \frac{1}{L_l} \left( f_{l,\textnormal{in}}^c(t) - f_{l,\textnormal{out}}^c(t) \right)
    \quad \forall c \in \{1,\dots,C\}, \label{hov2:eq:update}
  \end{equation}
  \end{linenomath}
  which is a slightly generalized form of a traditional multi-class CTM update.
  \item For all ordinary and destination links,
  \begin{linenomath}
  \begin{equation}
    f_{l,\textnormal{in}}^c(t) = \sum_{i=1}^{M_\nu} f_{il}^c(t), \label{hov2:eq:ord_in}
  \end{equation}
  \end{linenomath}
  where $\nu$ is the beginning node of link $l$.
  \item For all origin links,
  \begin{linenomath}
  \begin{equation}
    f_{l,\textnormal{in}}^c(t) = d_l^c(t). \label{hov2:eq:org_in}
  \end{equation}
  \end{linenomath}
  \item For all ordinary and origin links,
  \begin{linenomath}
  \begin{equation}
    f_{l,\textnormal{out}}^c(t) = \sum_{j=1}^{N_\nu} f_{lj}^c(t), \label{hov2:eq:ord_out}
  \end{equation}
  \end{linenomath}
    where $\nu$ is the ending node of link $l$.
  \item For all destination links,
  \begin{linenomath}
  \begin{equation}
    f_{l,\textnormal{out}}^c(t) = S_l^c(t). \label{hov2:eq:dest_out}
  \end{equation}
  \end{linenomath}
\end{itemize}

\subsection{Simulation algorithm}
\label{hov2:subsec_sim_algorithm}
\begin{enumerate}
  \item Initialization:
  \begin{linenomath}
    \begin{align*}
      \rho_l^c(0) \quad &:= \quad \rho_{l,0}^c \\
      t \quad &:= \quad 0,
    \end{align*}
  \end{linenomath}
    for all $l \in \Lcal$, $c \in \{1,\dots,C\}$.
  \item Perform all control inputs that have been (optionally) specified by the modeler.
    For our purposes, this includes:
    \begin{enumerate}
      \item For each managed lane link, modify the sending function of the link model in accordance with the friction effect model.
      As discussed in more detail in Part I \citep{wright_macroscopic_2019}, for the particular link model of Appendix~\ref{hov:app_linkmodel}, our friction effect model is to modify the sending function of a managed lane link from \eqref{hov:eq_lnctm_send_function} to
      \begin{linenomath}
        \begin{equation}
        S_{\MLnorm}^c(t) =
        \hat{v}_{\MLnorm}(t)\rho_{\MLnorm}^c(t)\min\left\{1,
        \frac{\hat{F}_{\MLnorm}(t)}{\hat{v}_{\MLnorm}(t)\sum_{c=1}^C \rho_{\MLnorm}^c(t)}\right\}
        \label{hov2:eq_friction_send_function}
        \end{equation}
        \end{linenomath}
      where the subscript ``ML'' means that the symbols refer to the parameters of the managed lane link, and where the friction-adjusted free flow speed $\hat{V}_\MLnorm(t)$ and capacity $\hat{F}_\MLnorm(t)$ are
      \begin{linenomath}
        \begin{align}
        \hat{v}_{\MLnorm}(t) & = v_{\MLnorm}^f(t) - \sigma_{\MLnorm}\Delta_{\MLnorm}(t)
        \label{hov2:eq_friction_free_flow_speed} \\
        \hat{F}_{\MLnorm}(t) & = \hat{v}_{\MLnorm}(t)\rho_{\MLnorm}^+ \label{hov2:eq_friction_capacity}
        \end{align}
      \end{linenomath}
      with $n_\MLnorm^+$ the high critical density (see Appendix \ref{hov:app_linkmodel} for the definition), $v^f_\MLnorm$ is the nominal free flow speed for the managed lane link, $\sigma_\MLnorm \in [0,1]$ is the \emph{friction coefficient} of the managed lane link (which quantifies the degree to which the friction effect exerts its influence on the vehicles in the link: \citet{jang_dual_2012} and others note that road configurations where the managed lane(s) and the GP lane(s) are more physically separated, e.g., a concrete barrier or traffic bollards instead of only a painted line, have a lower magnitude of the friction effect), and $\Delta_\MLnorm(t)$ is the speed differential between the managed lane link and the GP link,
      \begin{linenomath}
        \begin{equation}
        \Delta_{11}(t) = v_{11}^f - v_1(t-1).
        \label{hov2:eq_delta}
        \end{equation}
      \end{linenomath}
      \item For each managed lane link whose downstream node is a gate node in a gated-access managed lane configuration, perform class switching to ensure that vehicles realistically leave the managed lane to take a downstream offramp, as detailed in \citet{wright_macroscopic_2019}.
    \end{enumerate}
  \item For each link $l \in \Lcal$ and commodity $c \in \{1,\dots,C\}$, compute the demand, $S_l^c(t)$ using the link's link model.
  \item For each ordinary and destination link $l \in \Lcal$, compute the supply $R_l(t)$ using the link model.
  For origin links, the supply is not used.
  \item For each node $\nu \in \Ncal$ that has one or more undefined split ratios $\beta_{ij}^c(t)$, use the node's split ratio solver to complete a fully-defined set of split ratios.
    Note that if an inertia effect model is being used, the modified split ratio solver, e.g. the one described in Appendix \ref{hov:app_splitratiosolver}, should be used where appropriate.
  \item For each node $\nu \in \Ncal$, use the node model to compute throughflows $f_{ij}^c(t)$ for all $i,j,c$.
  \item For every link $l \in \Lcal$, compute the updated state $\vec{\rho}_l(t+1)$:
  \begin{itemize}
    \item If $l$ is an ordinary link, use \eqref{hov2:eq:update}, \eqref{hov2:eq:ord_in}, and \eqref{hov2:eq:ord_out}.
    \item If $l$ is an origin link, use \eqref{hov2:eq:update}, \eqref{hov2:eq:org_in}, and \eqref{hov2:eq:ord_out}
    \item If $l$ is a destination link, use \eqref{hov2:eq:update}, \eqref{hov2:eq:ord_in}, and \eqref{hov2:eq:dest_out}.
  \end{itemize}
  \item If $t=T$, then stop. Otherwise, increment $t:=t+1$ and return to step 2.
\end{enumerate}
 
\section{Calibrating the Managed Lane-Freeway Network Model}\label{hov2:sec_calibration}
Typically, a traffic modeler will have some set of data collected from traffic detectors (e.g., velocity and flow readings), and will create a network topology with parameter values that allow the model to reproduce these values in simulation.
Then, the parameters can be tweaked to perform prediction and analysis.
For our managed lane-freeway networks, the parameters of interest are:

\begin{enumerate}
\item Link model parameters for each link (also called ``fundamental diagram'' parameters, referring to the graph drawn by the sending and receiving functions as functions of density).
Calibration of a link model is typically agnostic to the node model and network topology,
and there exists an abundant literature on this topic.
For the purposes of the simulations in this work, we used the method of
\cite{dervisoglu_automatic_2009}, but any other method is appropriate.
\item Percentage of special (that is, able to access the managed lane) vehicles in the traffic flow entering the system.
This parameter depends on, e.g., the time of day and location as well as on the type
of managed lane.
It could be roughly estimated as a ratio of the managed lane vehicle count to the
total freeway vehicle count during periods of congestion at any given
location (supposing of course that congestion in the GP lanes will lead the eligible vehicles to select the managed lane to avoid this congestion).
\item Inertia coefficients.
These parameters affect only how traffic of different classes mixes
in different links, but they have no effect on the total vehicle counts 
produced by the simulation.
\item Friction coefficients.
How to tune these parameters is an open question.
In~\cite{jang_dual_2012} the dependency of a managed lane's speed
on the GP lane speed was investigated under different densities of
the managed lane, and the presented data suggests that although
the correlation between the two speeds exists, it is not overwhelmingly strong,
below 0.4.
Therefore, we suggest setting friction coefficients to values not exceeding
$0.4$.
\item Mutual restriction intervals for the partial FIFO constraint.
It is also an open question how to estimate mutual restriction intervals
from the measurement data.
See the discussion in \citet[Section 3.2]{wright_macroscopic_2019} for some guidelines.
Note also that the choice of restriction intervals govern the magnitude of the smoothing effect.
\item Offramp split ratios. \label{hov2:item:offrampsplits}
\end{enumerate}

Calibrating a traffic model, or identifying the best values of its parameters to match real-world data, is typically an involved process for all but the simplest network topologies.
In particular, once we consider more than a single, unbroken stretch of freeway, the nonlinear nature and network effects of these systems means that estimating each parameter in isolation might lead to unpredictable behavior.
Instead, nonlinear and/or non-convex optimization techniques such as genetic algorithms \citep{poole_metanet_2012}, particle swarm methods \citep{poole_second_2016}, and others (\citet{ngoduy_calibration_2012, fransson_framework_2012}, etc.) are employed.

In the managed lane-freeway networks we have discussed, the key unknown parameters we have introduced are the offramp split ratios, item \ref{hov2:item:offrampsplits}, which may be particularly hard to estimate as they are typically time-varying and explicitly represent driver behavior, rather than physical parameters of the road.
Estimating the values of the other parameters can be done with one of many methods in the literature.
The remainder of this section describes iterative methods for identification of the offramp split ratios for both the full-access and gated-access topological configurations, both of which were introduced in Part I \citep{wright_macroscopic_2019}.

\subsection{Split ratios for a full-access managed lane}\label{hov2:subsec_beta_fa}
\begin{figure}[htb]
    \centering
    \includegraphics[width=2in]{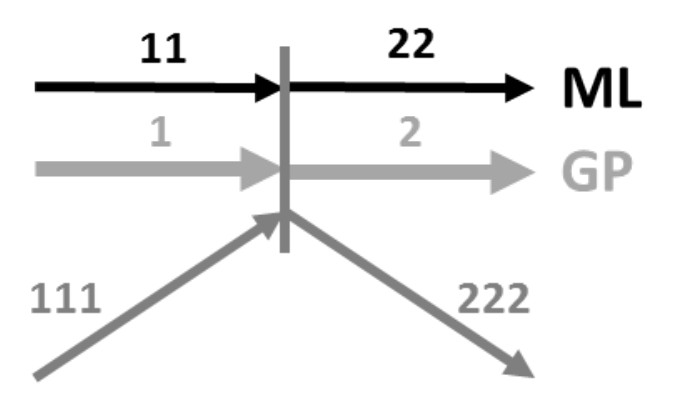}
    \caption[A node where some of the input links form travel facilities with some
    of the output links.]{A node where some of the input links form travel facilities with some
    of the output links. ML = Managed Lane.}
    \label{hov2:fig-node-1}
\end{figure}

Consider a node, one of whose output links is an offramp, as
depicted in Figure~\ref{hov2:fig-node-1}.
We shall make the following assumptions.
\begin{enumerate}
\item The total flow entering the offramp, $\hat{f}_{222}^{in}$, at any given time
is known (from measurements) and is not restricted by the offramp supply:
$\hat{f}_{222}^{in}<R_{222}$.
\item The portions of traffic sent to the offramp from the managed lane and from
the GP lane at any given time are equal:
$\beta_{1,222}^c = \beta_{11,222}^c \triangleq \beta$, $c=1,\dots,C$.
\item None of the flow coming from the onramp (link~111), if such flow
exists, is directed toward the offramp.
In other words, $\beta_{111,222}^c=0$, $c=1,\dots,C$.
\item The distribution of flow portions not directed to the offramp
between the managed lane and the GP output links is known.
This can be written as:
$\beta_{ij}^c = (1-\beta)\delta_{ij}^c$,
where $\delta_{ij}^c\in[0,1]$, as well as $\beta_{111,j}$,
$i=1,11$, $j=2,22$, $c=1,\dots,C$, are known.
\item The demand $S_i^c$, $i=1,11,111$, $c=1,\dots,C$, and supply $R_j$, $j=2,22$,
are given.
\end{enumerate}
At any given time, $\beta$ is unknown and is to be found.

If $\beta$ were known, the node model would compute the input-output flows,
in particular, $f_{i,222} = \sum_{c=1}^C f_{i,222}^c$, $i=1,11$.
Define
\begin{linenomath}
\be
\psi(\beta)=f_{1,222} + f_{11,222} - \hat{f}_{222}^{in}.
\label{hov2:psi_def}
\ee
Our goal is to find $\beta$ from the equation
\be
\psi(\beta) = 0,
\label{hov2:eq_psi_beta}
\ee
\end{linenomath}
such that 
$\beta\in\left[\frac{\hat{f}_{222}^{in}}{S_1+S_{11}},1\right]$,
where $S_i=\sum_{c=1}^CS_i^c$.
Obviously, if $S_1+S_{11} < \hat{f}_{222}^{in}$, the solution does not exist,
and the best we can do in this case to match $\hat{f}_{222}^{in}$ is to set $\beta=1$, directing
all traffic from links~1 and~11 to the offramp.

Suppose now that $S_1+S_{11} \geq \hat{f}_{222}^{in}$.
For any given $\hat{f}_{222}^{in}$, we assume $\psi(\beta)$
is a monotonically increasing function of $\beta$ (this assumption is true for the particular node model of \citet{wright_node_2017}).
Moreover, $\psi\left(\frac{\hat{f}_{222}^{in}}{S_1+S_{11}}\right)\leq 0$,
while $\psi(1)\geq 0$.
Thus, the solution of~\eqref{hov2:eq_psi_beta} within the given interval exists
and can be obtained using the \emph{bisection method}.

The algorithm for finding $\beta$ follows.
\begin{enumerate}
\item Initialize:
\begin{eqnarray*}
\underline{b}(0) & := & \frac{\hat{f}_{222}^{in}}{S_1+S_{11}};\\
\overline{b}(0) & := & 1;\\
k & := & 0.
\end{eqnarray*}
\item If $S_1+S_{11} \leq \hat{f}_{222}^{in}$, then are not enough vehicles to satisfy the offramp demand. Set $\beta = 1$ and stop.
\item Use the node model with
$\beta=\underline{b}(0)$ and evaluate $\psi(\beta)$.
If $\psi(\underline{b}(0))\geq 0$, then set $\beta=\underline{b}(0)$ and stop.
\item Use the node model with
$\beta=\frac{\underline{b}(k)+\overline{b}(k)}{2}$ and evaluate $\psi(\beta)$.
If $\psi\left(\frac{\underline{b}(k)+\overline{b}(k)}{2}\right)=0$,
then set $\beta=\frac{\underline{b}(k)+\overline{b}(k)}{2}$ and stop.
\item If $\psi\left(\frac{\underline{b}(k)+\overline{b}(k)}{2}\right) < 0$,
then update:
\begin{eqnarray*}
\underline{b}(k+1) & = & \frac{\underline{b}(k)+\overline{b}(k)}{2}; \\
\overline{b}(k+1) & = & \overline{b}(k).
\end{eqnarray*}
Else, update:
\begin{eqnarray*}
\underline{b}(k+1) & = & \underline{b}(k);\\
\overline{b}(k+1) & = & \frac{\underline{b}(k)+\overline{b}(k)}{2}.
\end{eqnarray*}
\item Set $k:=k+1$ and return to step 4.
\end{enumerate}

Here, $\underline{b}(k)$ represents the lower bound of the search interval at iteration $k$ and $\overline{b}(k)$ the upper bound.

\subsection{Split ratios for the separated managed lane with gated access}\label{hov2:subsec_beta_sep}
\begin{figure}[htb]
\centering
\includegraphics[width=2in]{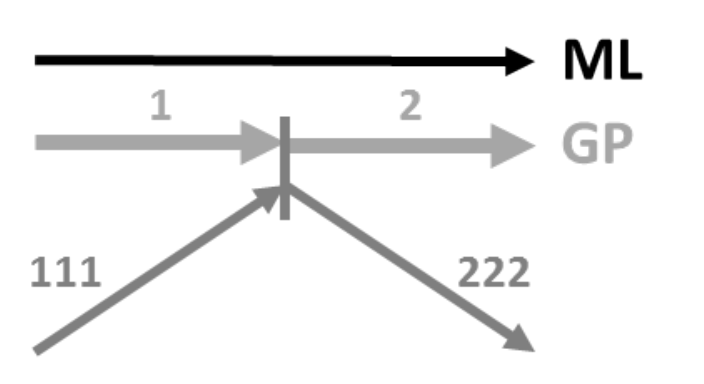}
\caption{A node with a GP link and an onramp as inputs, and 
a GP link and an offramp as outputs.}
\label{hov2:fig-b-sep}
\end{figure}

The configuration of a node with an offramp as one of the output links
is simpler in the case of a separated managed lane, as shown in
Figure~\ref{hov2:fig-b-sep}.
Here, traffic cannot directly go from the managed lane to link~222,
and, thus, we have to deal only with the 2-input-2-output node.
There is a caveat, however.
Recall from the discussion of gated-access managed lanes in \citet{wright_macroscopic_2019}
that in the separated managed lane case we have destination-based 
traffic classes, and
split ratios for destination-based traffic are fixed due to being determined at an outer loop level.

We shall make the following assumptions:
\begin{enumerate}
\item The total flow entering the offramp, $\hat{f}_{222}^{in}$, at any given time
is known (from measurements) and is not restricted by the offramp supply:
$\hat{f}_{222}^{in}<R_{222}$.
\item All the flow coming from the onramp (link~111), if such flow
exists, is directed toward the GP link~2.
In other words, $\beta_{111,2}^c=1$ and $\beta_{111,222}^c=0$, $c=1,\dots,C$.
\item The demand $S_i^c$, $i=1,111$, $c=1,\dots,C$, and supply $R_2$ are given.
\item We denote the set of destination-based classes as $\DD$.
The split ratios $\beta_{1j}^c$ for $c\in\DD$ are known.
Let the split ratios $\beta_{1j}^c = \beta$ for $c\in \{1,\dots,C\} \setminus \DD$, where $\beta$ is to be determined (i.e., we assume all non-destination-based classes exit at the same rate).
\end{enumerate}
The first three assumptions here reproduce assumptions~1,~3 and~5
made for the full-access managed lane case.
Assumption~4 is a reminder that there is a portion of traffic flow that we
cannot direct to or away from the offramp, but we have to account for it.

Similarly to the full-access managed lane case, we define the function $\psi(\beta)$:
\begin{linenomath}
\be
\psi(\beta) = \sum_{c\in\overline{\DD}}f_{1,222}^c +
\sum_{c\in\DD}f_{1,222}^c - \hat{f}_{222}^{in},
\label{hov2:def-psi2}
\ee
\end{linenomath}
where $f_{1,222}^c$, $c=1,\dots,C$ are determined by the node model.
The first term of the right-hand side of~\eqref{hov2:def-psi2} depends on
$\beta$.
As before, we assume $\psi(\beta)$ is a monotonically increasing function.
We look for the solution of equation~\eqref{hov2:eq_psi_beta} on the interval
$[0,1]$.
This solution exists iff $\psi(0)\leq 0$ and $\psi(1)\geq 0$.
The algorithm for finding $\beta$ is the same as the one presented
in the previous section, except that $\underline{b}(0)$ should be
initialized to $0$, and $S_{11}$ is to be assumed $0$.

\subsection{An iterative full calibration process}\label{hov2:subsec-calibration-process}
\begin{figure}[htb]
\centering
\includegraphics[width=\textwidth]{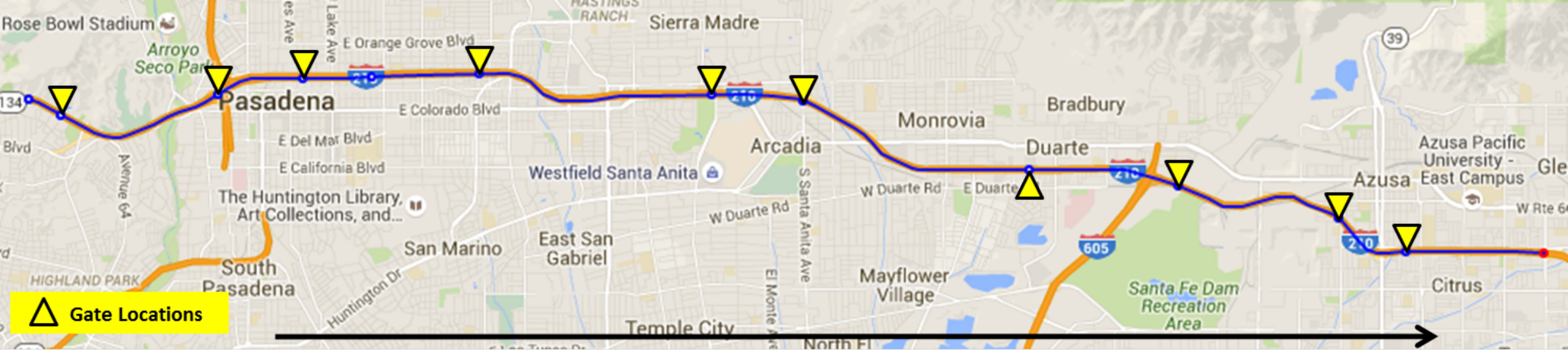}
\caption{Calibration workflow.}
\label{hov2:fig-calibration}
\end{figure}

For the purposes of the simulations presented in the following Section, we placed the iterative split ratio identification methods of Sections \ref{hov2:subsec_beta_fa} and \ref{hov2:subsec_beta_sep} within a larger iterative loop for the remaining parameters.
The model calibration follows the flowchart shown in
Figure~\ref{hov2:fig-calibration}.

\begin{enumerate}
\item We start by assembling the available measurement data.
Fundamental diagrams are assumed to be given.
Mainline and onramp demand are specified per 5-minute periods together
with the special vehicle portion parameter indicating the fraction of the input demand
that is able to access the managed lane.
Initially, we do not know offramp split ratios as they cannot be measured
directly.
Instead, we use some arbitrary values to represent them and call these values ``initially
guessed offramp split ratios''.
Instead of the offramp split ratios, we have the measured flows actually observed on the offramps, which we refer to as \emph{offramp demand}.
\item We run our network simulation outlined in Section \ref{hov2:subsec_sim_algorithm} for the entire simulation period.
At this point, in step~5 of the simulation, the \emph{a priori} undefined split ratios between traffic in the
GP and in the managed lanes are assigned using a split ratio solver.
\item Using these newly-assigned split ratios, we run our network simulation
again, only this time, instead of using the initially guessed offramp split ratios, we
compute them from the given offramp demand as described in 
Sections~\ref{hov2:subsec_beta_fa} and~\ref{hov2:subsec_beta_sep}.
As a result of this step, we obtain new offramp split ratios.
\item Now we run the network simulation as we did originally, in step~2,
only this time with new offramp split ratios, and record the simulation
results --- density, flow, speed, as well as performance measures such
as vehicle miles traveled (VMT) and vehicle hours traveled (VHT).
\item Check if the resulting offramp flows match the offramp demand.
If yes, proceed to step~6, otherwise, repeat steps~2-5.
In our experience (i.e., the case studies in the following Section), it takes the process described in steps~2-5 no more than two
iterations to converge.
\item Evaluate the simulation results:
\begin{itemize}
\item correctness of bottleneck locations and activation times;
\item correctness of congestion extension at each bottleneck;
\item correctness of VMT and VHT.
\end{itemize}
If the simulation results are satisfactory, stop.
Otherwise, proceed to step~7.
\item Tune/correct input data in the order shown in block~7 of
Figure~\ref{hov2:fig-calibration}.
\end{enumerate}

\section{Simulation Results}\label{hov2:sec_simulation}
\subsection{Full-access managed lane case study: Interstate 680 North}\label{hov2:subsec_i680}
\begin{figure}[htb]
\centering
\includegraphics[width=\textwidth]{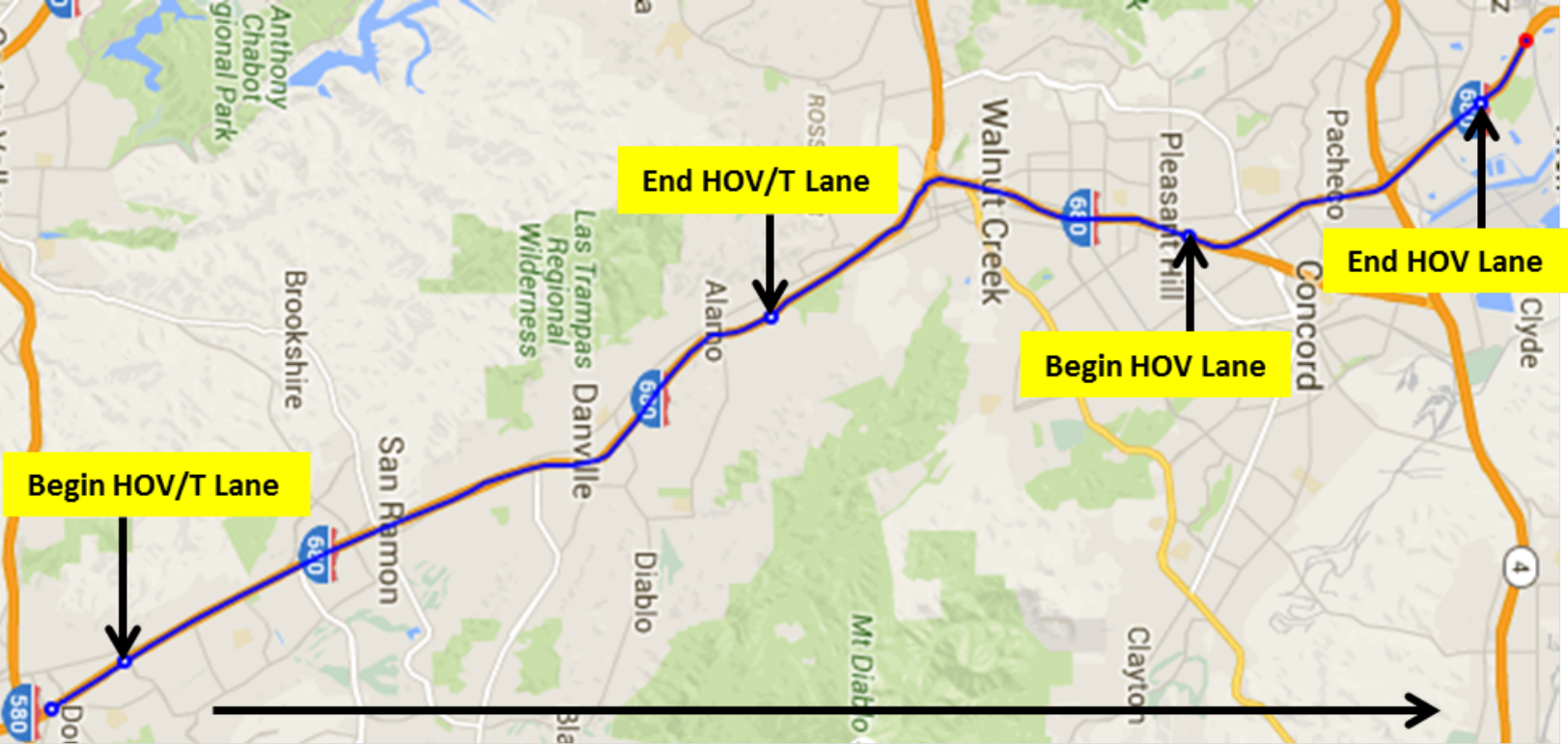}
\caption{Map of I-680 North in Contra Costa County.}
\label{hov2:fig-680-map}
\end{figure}

We consider a 26.8-mile stretch of I-680 North freeway in Contra Costa County, California,
from postmile 30 to postmile 56.8, shown in Figure~\ref{hov2:fig-680-map},
as a test case for the full-access managed lane configuration. 
This freeway's managed lane facilities are split into two segments whose beginning and end points are marked on the map. 
The first segment is a high-occupancy-tolled (HOT) lane, which allows free entry to vehicles with two or more passengers and tolled entry to single-occupancy vehicles \citep{metropolitan_transportation_commission_bay_2019},
The second segment has an HOV lane is 4.5 miles long.
There are 26 onramps and 24 offramps.
The HOV lane is active from 5 to 9 AM and from 3 to 7 PM.
The rest of the time, the HOV lane is open to all traffic, and behaves as a GP lane.

To build the model, we used data collected for the
I-680 Corridor System Management Plan (CSMP) study~\citep{system_metrics_group_inc._contra_2015}.
The bottleneck locations as well as their activation times and congestion
extension were identified in that study using video
monitoring and tachometer vehicle runs.
On- and offramp flows were given in 5-minute increments.
For the purposes of our model, we do not consider tolling dynamics, and instead assume that managed lane-eligible vehicles incur no cost to access the managed lane. 
We assume that the managed lane-eligible portion of the input demand is 15\%.
The model was calibrated to a typical weekday, as suggested
in the I-680 CSMP study.

For this simulation, we used the fundamental diagram described in Appendix~\ref{hov:app_linkmodel}, with parameters as follows:
\begin{itemize}
\item The capacity of the ordinary GP lane is 1,900 vehicles per hour per lane (vphl);
\item The capacity of the auxiliary GP lane is 1,900 vphl;
\item The capacity of the managed lane is 1,800 vphl while active and 1,900 vphl
when it behaves as a GP lane;
\item The free flow speed varies between 63 and 70 mph --- these measurements
came partially from the California Performance Measurement System (PeMS)~\citep{pems_california_2019} and partially from tachometer vehicle
runs.
\item The congestion wave speed for each link was taken as $1/5$ of the
free flow speed.
\end{itemize}

\begin{figure}[htb]
\centering
\includegraphics[width=\textwidth]{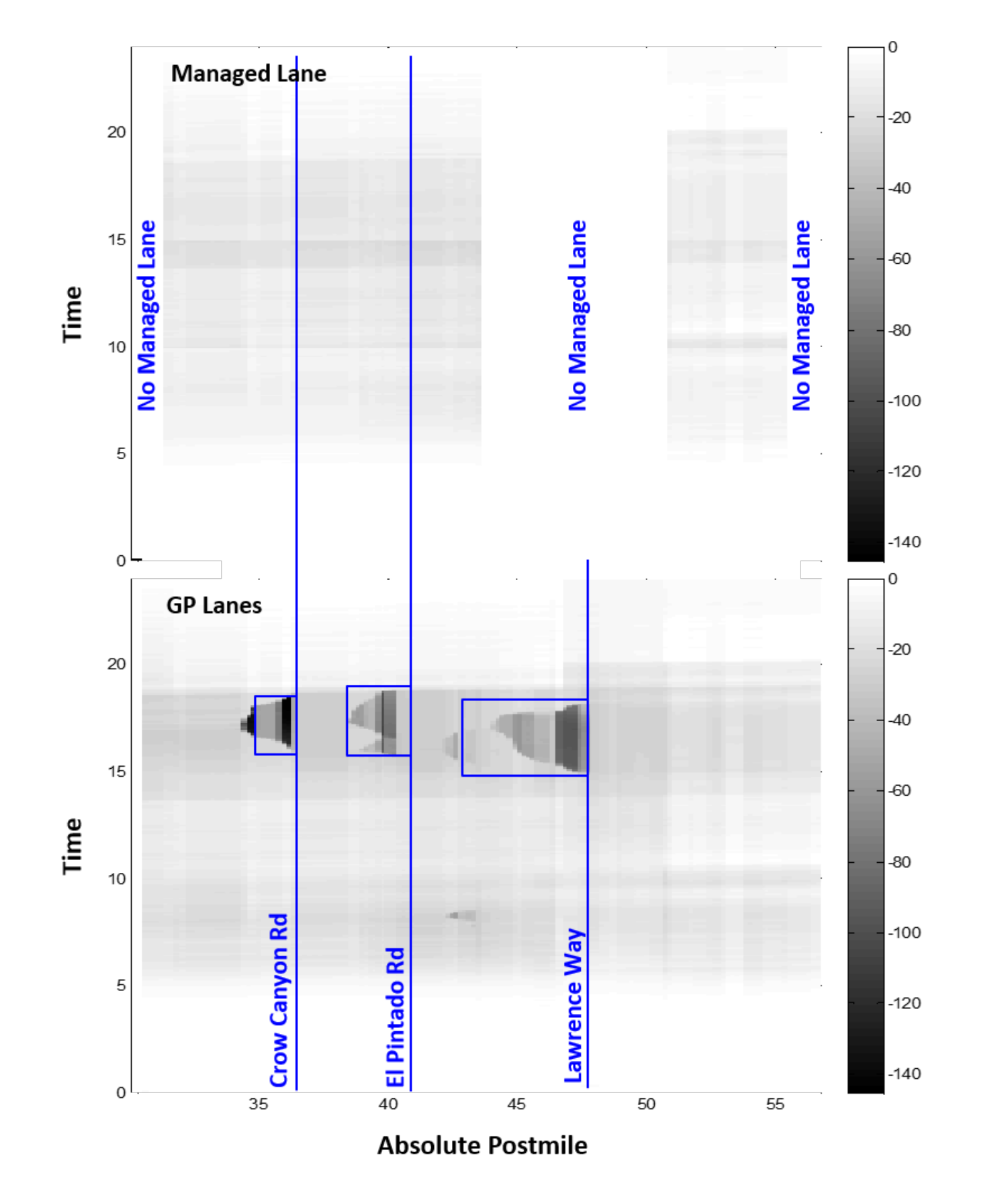}
\caption[I-680 North density contours for GP and managed lanes produced
by simulation.
Density values are given in vehicles per mile per lane.
]{I-680 North density contours for GP and managed lanes produced
by simulation.
Density values are given in vehicles per mile per lane.
Blue boxes on the GP lane speed contour indicate congested areas as identified
by the I-680 CSMP study.}
\label{hov2:fig-680-density}
\end{figure}

\begin{figure}[htb]
\centering
\includegraphics[width=\textwidth]{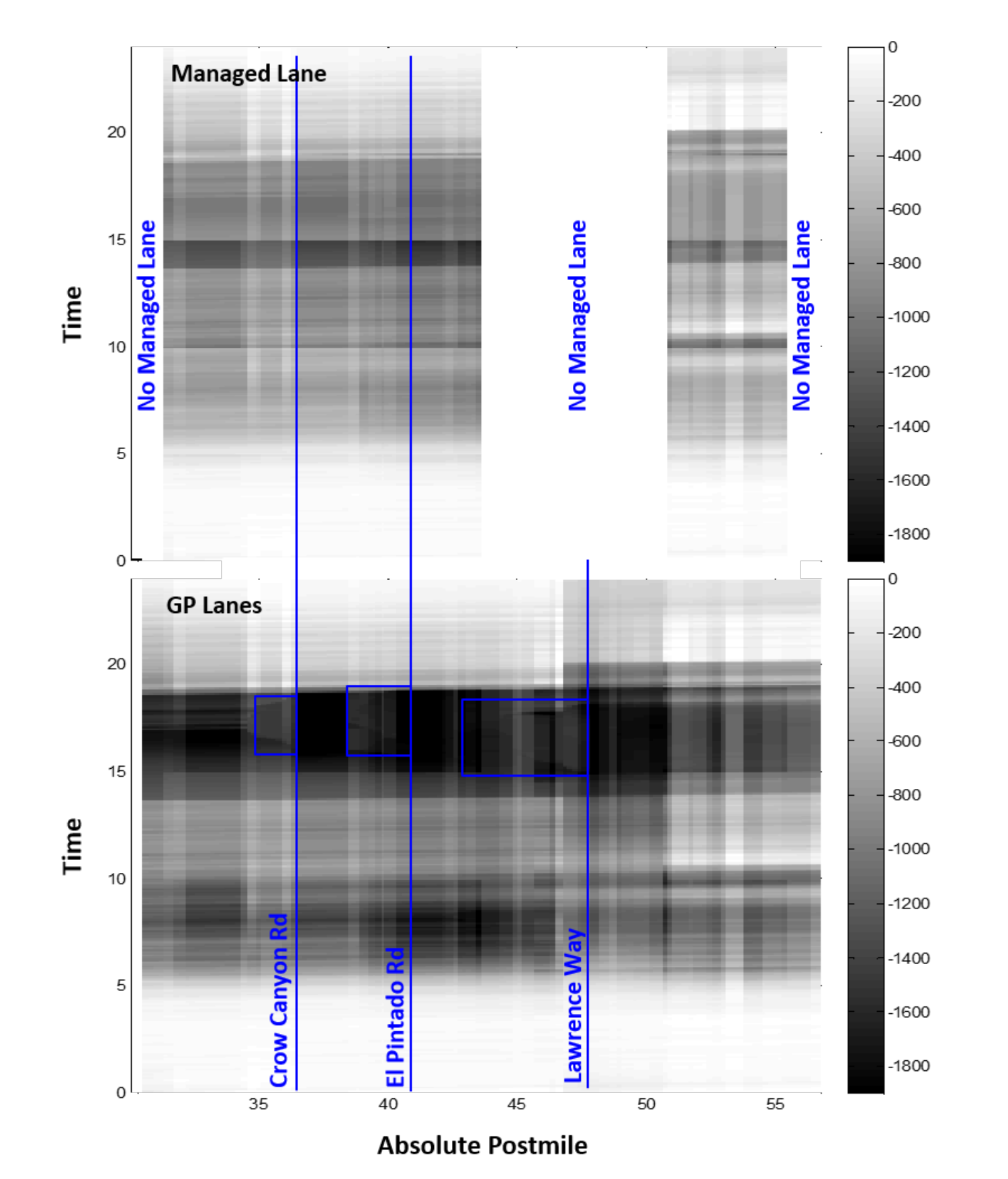}
\caption[I-680 North flow contours for GP and managed lanes produced 
by simulation.
Flow values are given in vehicles per hour per lane.
]{I-680 North flow contours for GP and managed lanes produced 
by simulation.
Flow values are given in vehicles per hour per lane.
Blue boxes on the GP lane speed contour indicate congested areas as identified
by the I-680 CSMP study.}
\label{hov2:fig-680-flow}
\end{figure}

\begin{figure}[htb]
\centering
\includegraphics[width=\textwidth]{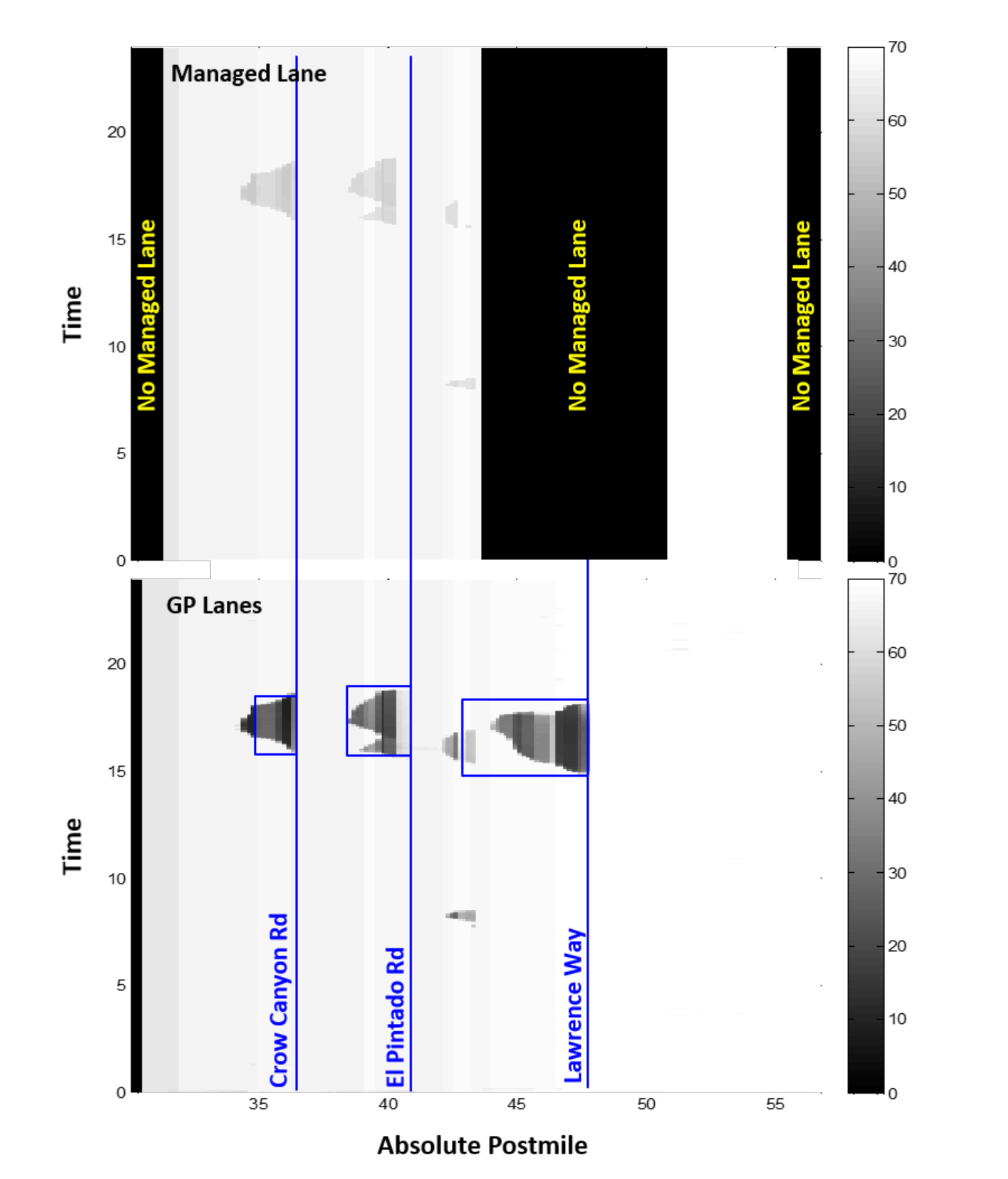}
\caption[I-680 North speed contours for GP and managed lanes produced by simulation.
Speed values are given in miles per hour.]{I-680 North speed contours for GP and managed lanes produced by simulation. 
Speed values are given in miles per hour.
Blue boxes on the GP lane speed contour indicate congested areas as identified
by the I-680 CSMP study.}
\label{hov2:fig-680-speed}
\end{figure}

\begin{figure}[htb]
\centering
\includegraphics[width=\textwidth]{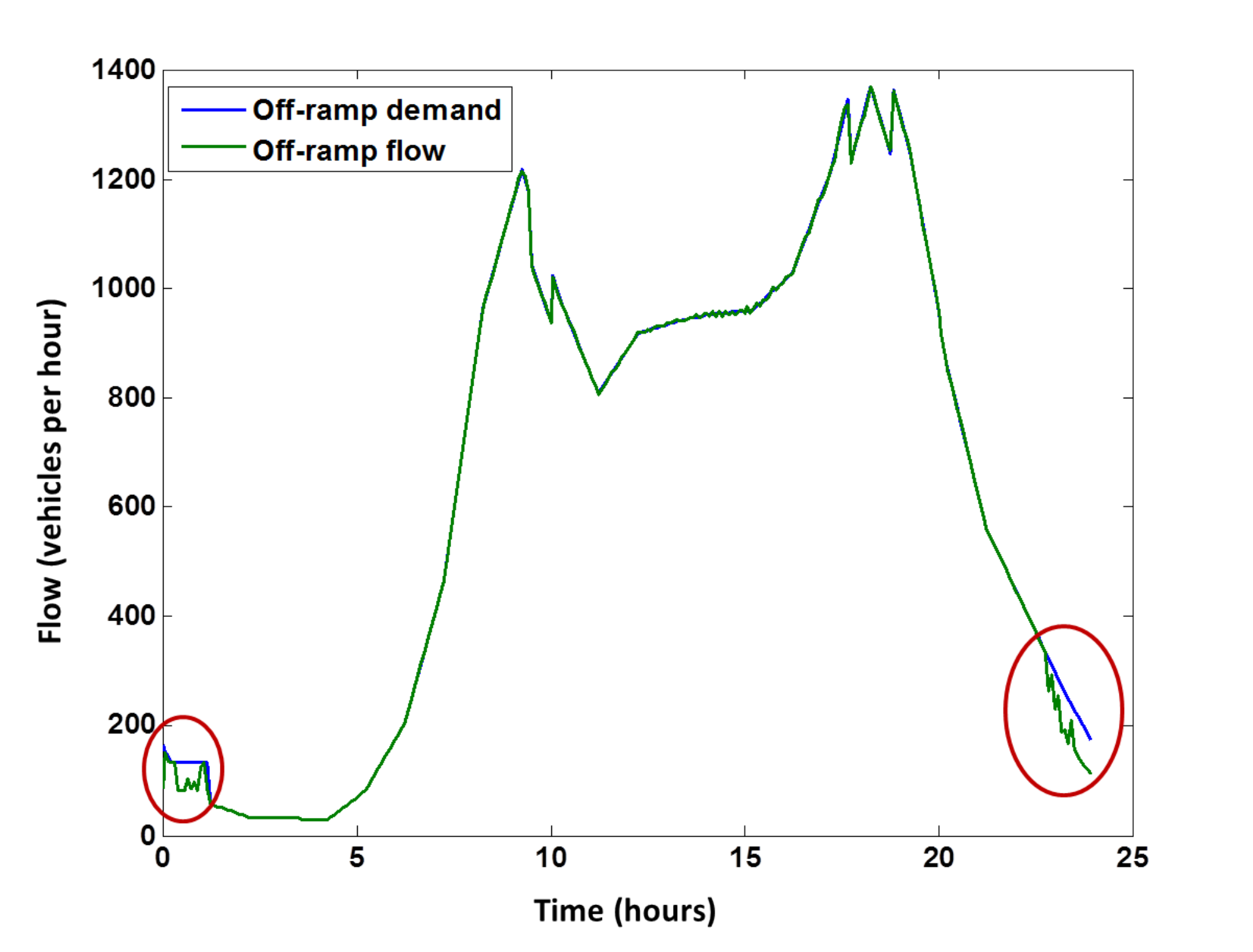}
\caption[Flow at the Crow Canyon Road offramp over 25 hours]{Flow at the Crow Canyon Road offramp over 24 hours --- collected
(offramp demand) vs. computed by simulation (offramp flow).}
\label{hov2:fig-680-offramp-flow}
\end{figure}

The modeling results are presented in Figures~\ref{hov2:fig-680-density},
\ref{hov2:fig-680-flow} and~\ref{hov2:fig-680-speed} showing density, flow and speed
contours, respectively, in the GP and the managed lanes.
In each plot, the top contour corresponds to the managed lanes,
and the bottom to the GP lanes.
In all the plots traffic moves from left to right along the
``Absolute Postmile'' axis, while the vertical axis represents time.
Bottleneck locations and congestion areas identified by the I-680 CSMP study
are marked by blue boxes in GP lane contours.
The managed lane does not get congested, but there is a speed drop due to the
friction effect.
The friction effect, when vehicles in the managed lane slow down because of the
slow moving GP lane traffic, can be seen in the managed lane speed contour
in Figure~\ref{hov2:fig-680-speed}.

Figure~\ref{hov2:fig-680-offramp-flow} shows an example of how well the
offramp flow computed by the simulation matches the target,
referred to as \emph{offramp demand}, as recorded by the detector on the offramp
at Crow Canyon Road.
We can see that in the beginning and in the end of the day, the computed flow
falls below the target (corresponding areas are marked with red circles).
This is due to the shortage of the mainline traffic in the simulation --- the offramp demand
cannot be satisfied.

Finally, Table~\ref{hov2:tab-680} summarizes the performance metrics ---
vehicle miles traveled (VMT), vehicle hours traveled (VHT) and delay
in vehicle-hours --- computed by simulation versus those collected in the course
of the I-680 CSMP study.
Delay is computed for vehicles with speed below 45 mph.

\begin{table}[ht]
    \centering
    \begin{tabular}{l r r }
        \toprule
    Performance Metric & Simulation result & Collected data  \\ \midrule
    GP Lane VMT & 1,687,618 & - \\ 
    Managed Lane VMT & 206,532 & - \\ 
    Total VMT & 1,894,150 & 1,888,885  \\ 
    GP Lane VHT & 27,732  & - \\ 
    Managed Lane VHT & 3,051 & - \\ 
    Total VHT & 30,783 & 31,008 \\ 
    GP Lane Delay (hr) & 2,785 & - \\ 
    Managed Lane Delay (hr) & 6 & - \\ 
    Total Delay (hr) & 2,791 & 2,904 \\ \bottomrule
    \end{tabular}
    \caption{Performance metrics for I-680 North.}
    \label{hov2:tab-680}
    \end{table}

\FloatBarrier

\subsection{Gated-access managed lane case study: Interstate 210 East}\label{hov2:subsec_i210}
\begin{figure}[htb]
\centering
\includegraphics[width=\textwidth]{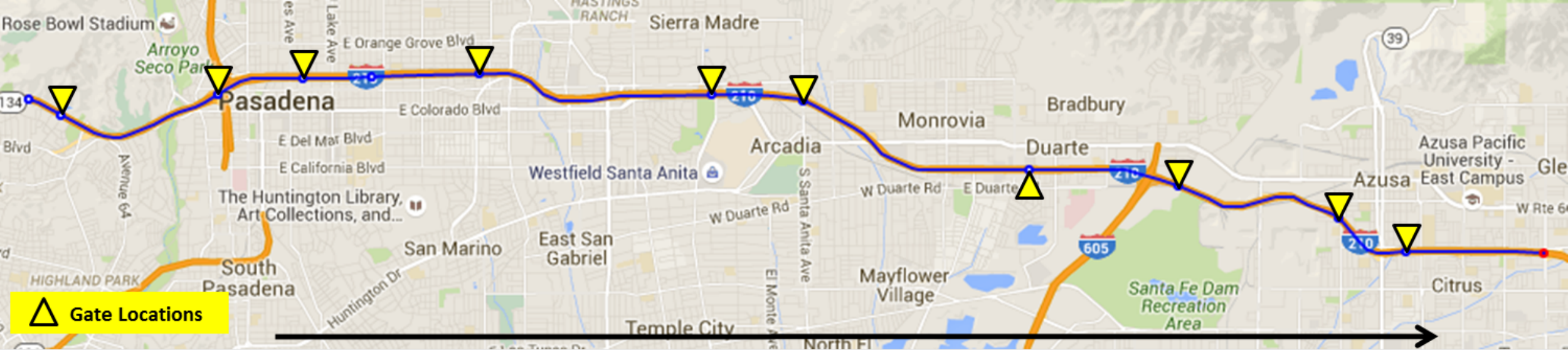}
\caption{Map of SR-134 East/ I-210 East freeway in Los Angeles County.}
\label{hov2:fig-210-map}
\end{figure}

We consider a 20.6-mile stretch of SR-134 East/ I-210 East in
Los Angeles County, California, shown in Figure~\ref{hov2:fig-210-map},
as a test case for the separated managed lane configuration.
This freeway's managed lane is also an HOV lane.
This freeway stretch consists of 3.9 miles of SR-134 East 
from postmile 9.46 to postmile 13.36, which merges into 16.7 miles of I-210 East
from postmile 25 to postmile 41.7.
Gate locations where traffic can switch between the GP and the HOV lanes
are marked on the map.
At this site, the HOV lane is always active.
There are 28 onramps and 25 offramps.
The largest number of offramps between two gates is 5.
Thus, our freeway model has 7 vehicle classes - LOV (low-occupancy vehicles; not managed lane-eligible), HOV (managed lane-eligible) and 5
destination-based.

To build the model, we used PeMS data for the corresponding segments
of the SR-134 East and I-210 East for Monday, October 13, 2014~\citep{pems_california_2019}.
Fundamental diagrams were calibrated using PeMS data
following the methodology of \citet{dervisoglu_automatic_2009}.
As in the I-680 North example, 
we assume that the managed lane-eligible portion of the input demand is 15\%.

\begin{figure}[htb]
\centering
\includegraphics[width=\textwidth]{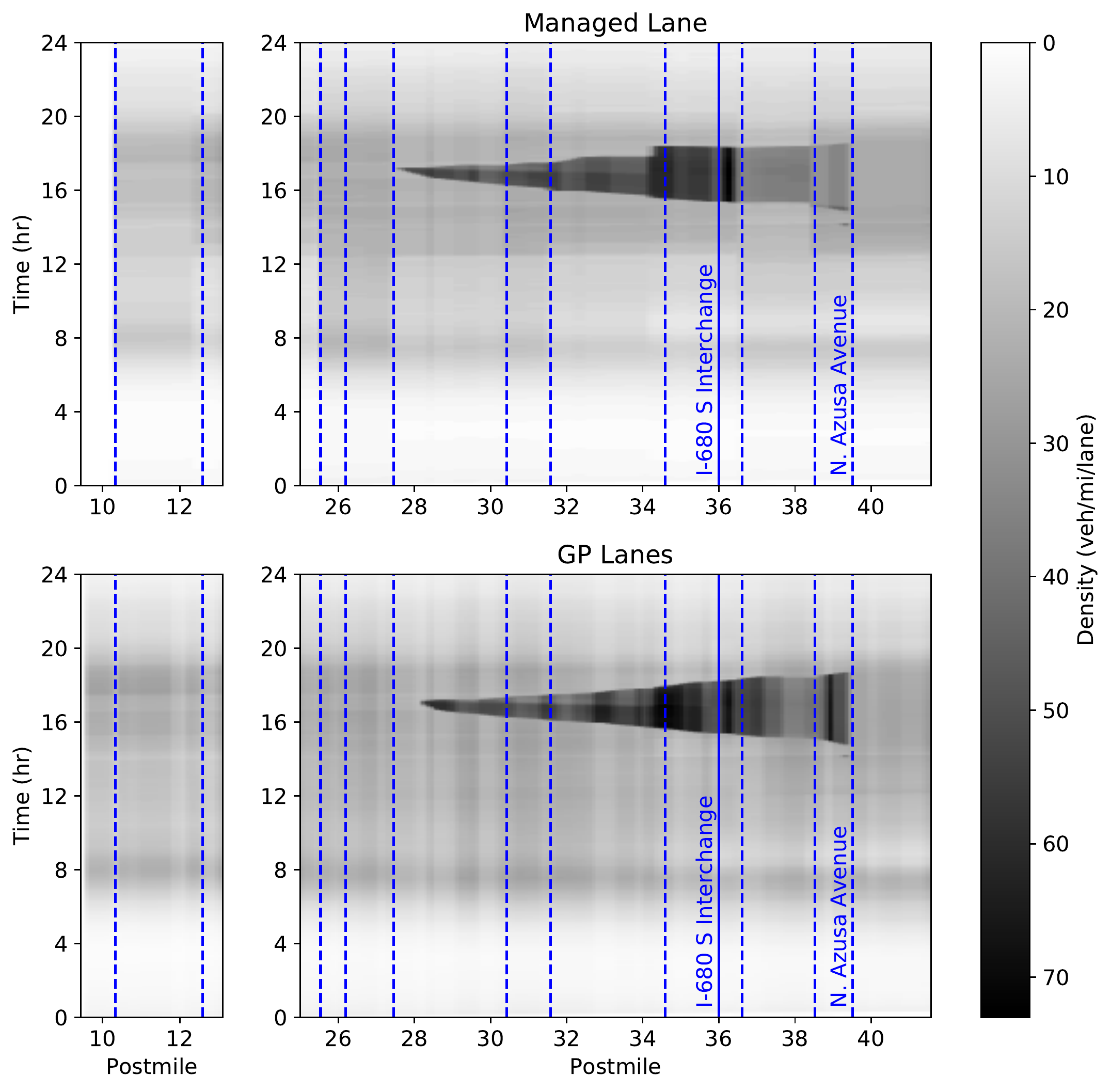}
\caption[SR-134 East/ I-210 East density contours for GP and managed lanes produced by simulation.]{SR-134 East (left plots) / I-210 East (right plots) density contours for GP and managed lanes produced
by simulation. Dotted blue lines represent the approximate position of the managed lane gates.}
\label{hov2:fig-210-density}
\end{figure}

\begin{figure}[htb]
\centering
\includegraphics[width=\textwidth]{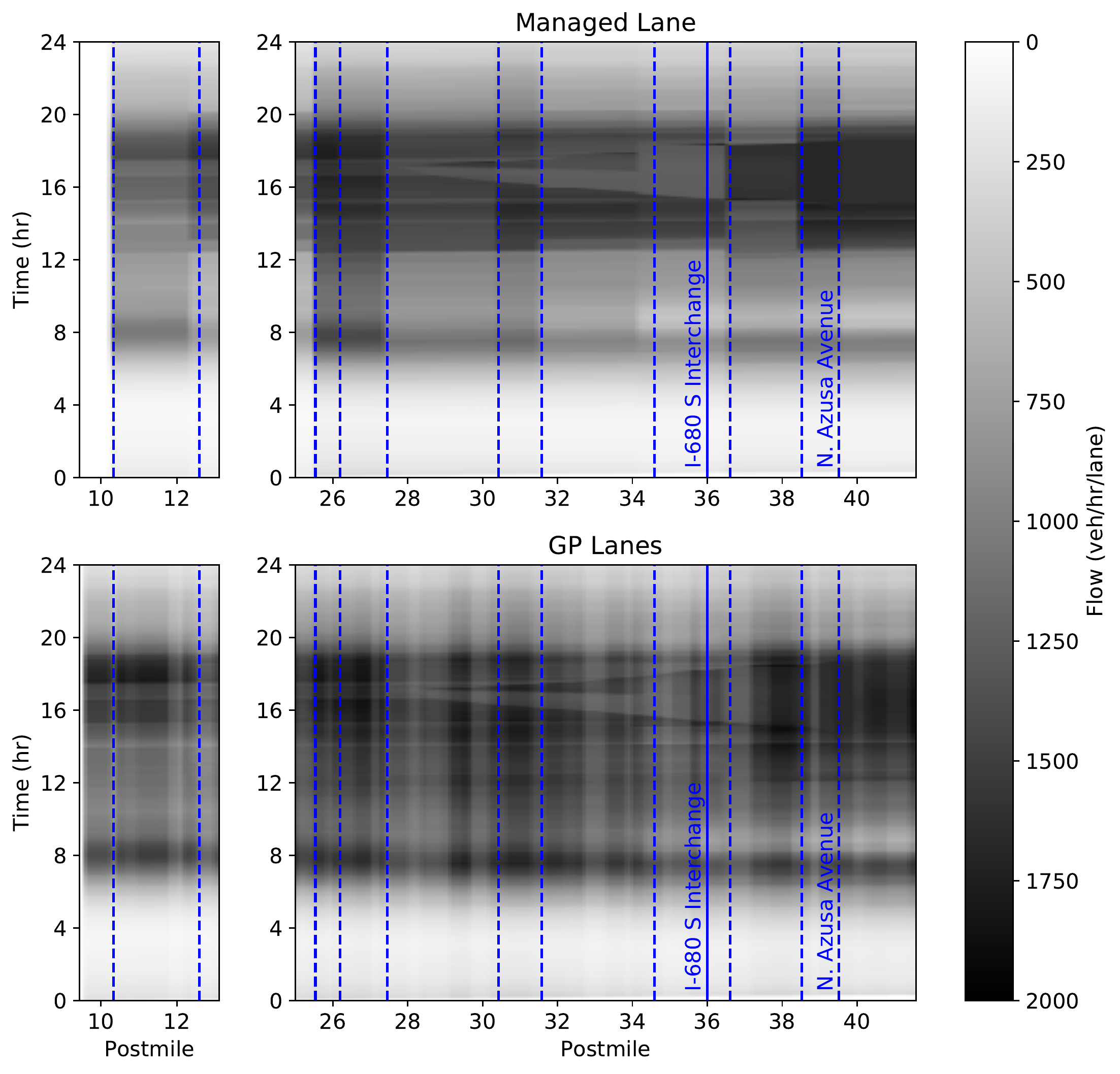}
\caption[SR-134 East/ I-210 East flow contours for GP and managed lanes produced 
by simulation.]{SR-134 East (left plots) / I-210 East (right plots) flow contours for GP and managed lanes produced 
by simulation. Dotted blue lines represent the approximate position of the managed lane gates.}
\label{hov2:fig-210-flow}
\end{figure}

\begin{figure}[htb]
\centering
\includegraphics[width=\textwidth]{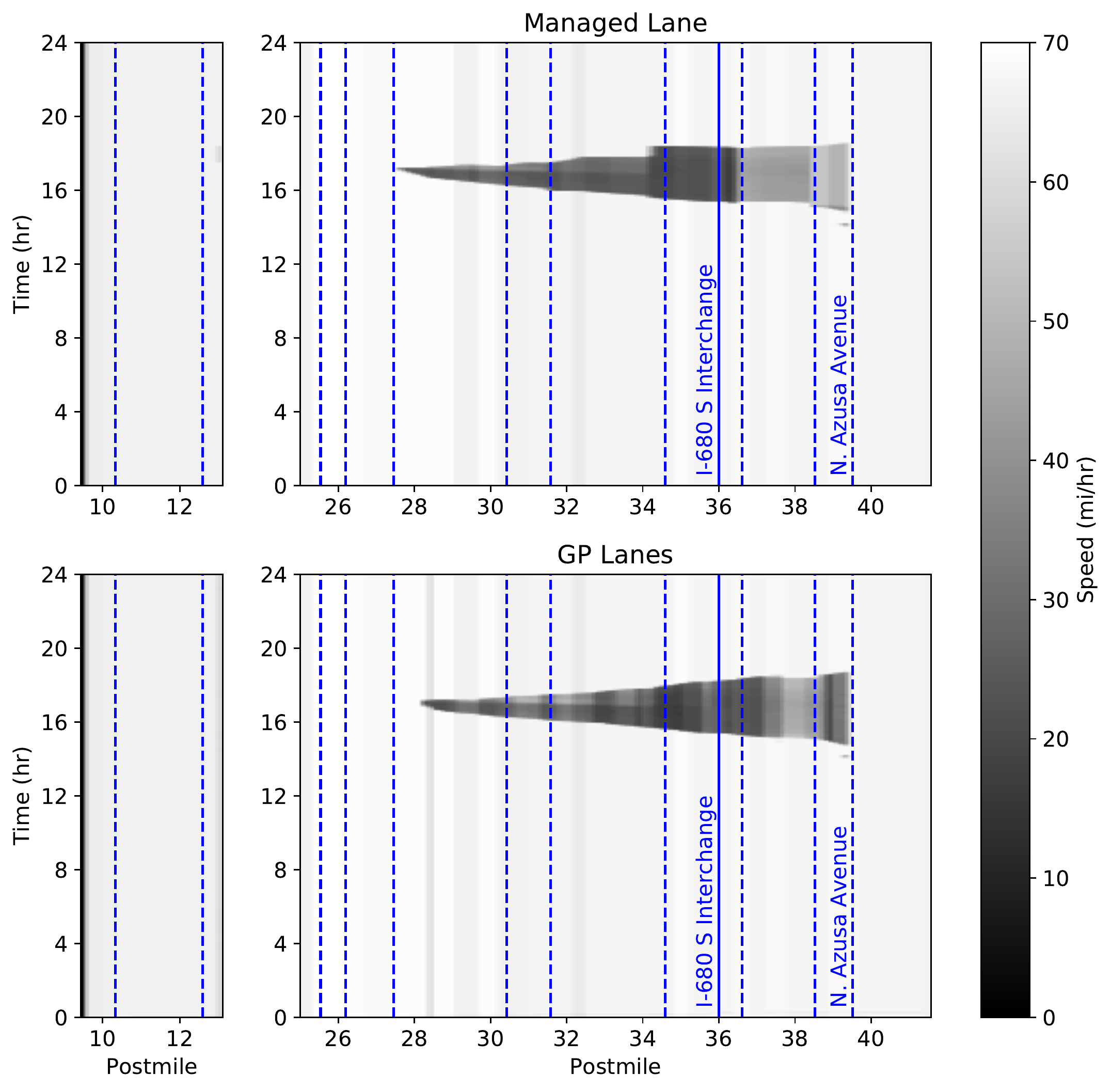}
\caption[SR-134 East/ I-210 East speed contours for GP and managed lanes produced 
by simulation.]{SR-134 East (left plots) / I-210 East (right plots) speed contours for GP and managed lanes produced 
by simulation. Dotted blue lines represent the approximate position of the managed lane gates.}
\label{hov2:fig-210-speed}
\end{figure}

\begin{figure}[htb]
\centering
\includegraphics[width=\textwidth]{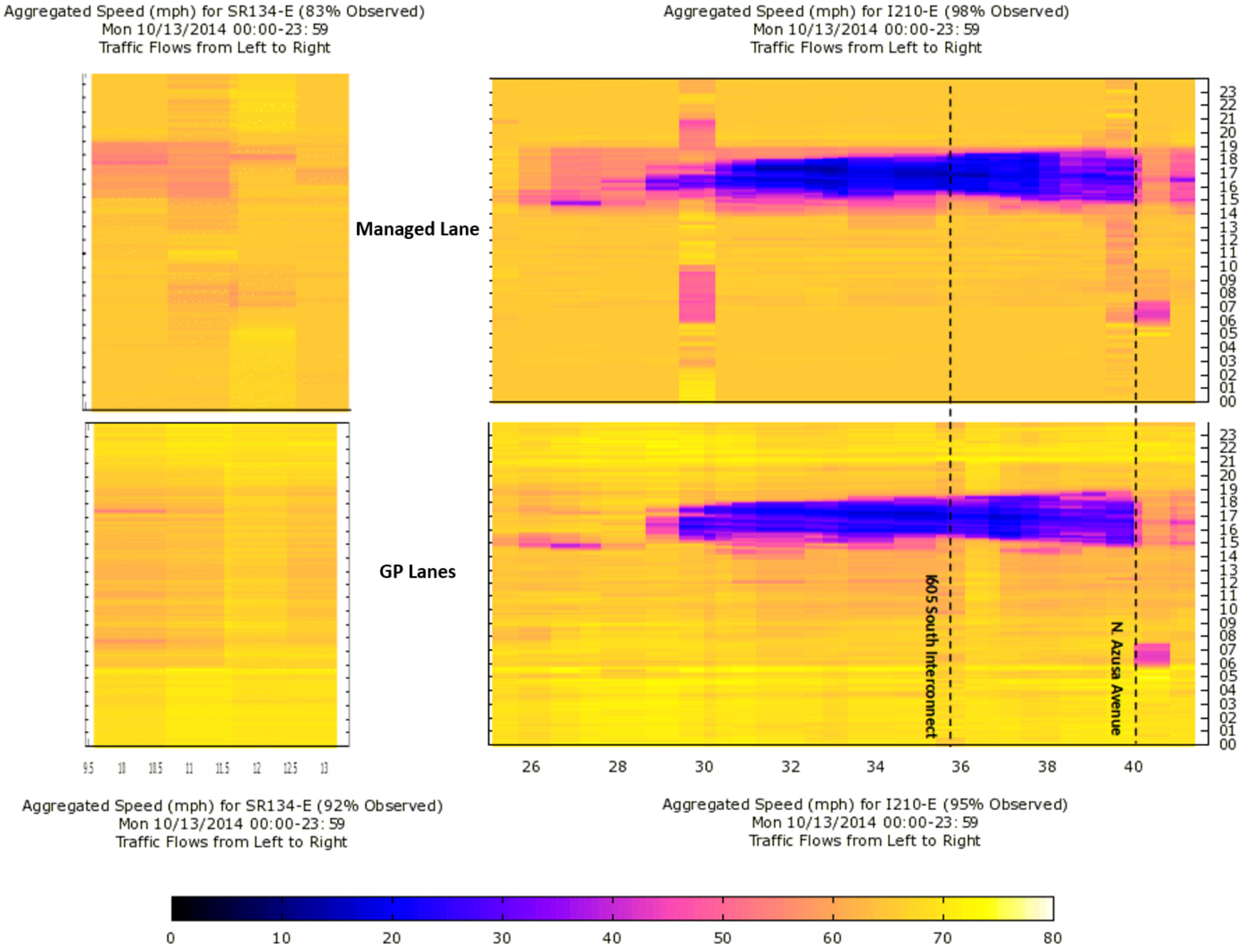}
\caption[SR-134 East/ I-210 East speed contours for GP and managed lanes obtained
from PeMS for Monday, October 13, 2014.]{SR-134 East/ I-210 East speed contours for GP and managed lanes obtained
from PeMS for Monday, October 13, 2014.
The horizontal axis represents absolute postmile, and the vertical axis represents
time in hours. Note the four contours share the same color scale.}
\label{hov2:fig-210-pems-speed}
\end{figure}

\begin{figure}[htb]
\centering
\includegraphics[width=\textwidth]{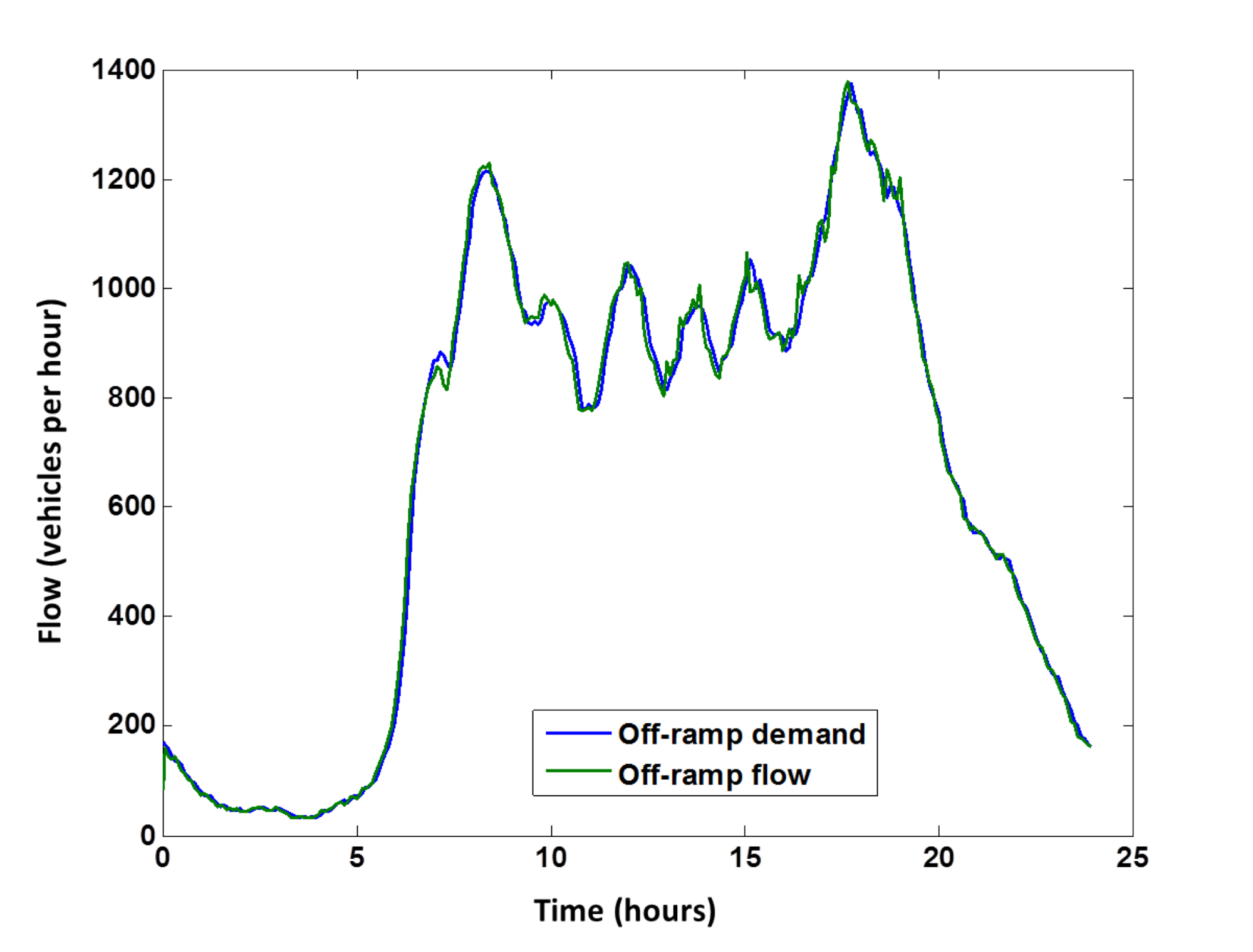}
\caption{Flow at the North Hill Avenue offramp over 24 hours --- PeMS data
(offramp demand) vs. computed by simulation (offramp flow).}
\label{hov2:fig-210-offramp-flow}
\end{figure}

The modeling results are presented in Figures~\ref{hov2:fig-210-density},
\ref{hov2:fig-210-flow} and~\ref{hov2:fig-210-speed} showing density, flow and speed
contours, respectively, in the GP and the managed lanes.
In each plot, the top contour corresponds to the managed lanes,
and the bottom to the GP lanes.
As before, in all the plots traffic moves from left to right along the
``Absolute Postmile'' axis, while the vertical axis represents time.
The managed lane does not get congested.
Dashed blue lines on the contour plots indicate managed gate locations.

Figure~\ref{hov2:fig-210-pems-speed} shows the PeMS speed contours for
the SR-134 East/ I-210 East GP and managed lanes that were used as a target
for our simulation model.
In these plots, traffic also travels from left to right, with the
horizontal axis representing postmiles, 
while the vertical axis represents time.

Figure~\ref{hov2:fig-210-offramp-flow} shows an example of how well the
offramp flow computed by the simulation matches the target,
referred to as \emph{offramp demand}, as recorded by the detector on the offramp
at North Hill Avenue.
The simulated offramp flow matches the offramp demand fairly closely.
Similar results were found for the other offramps.

Finally, Table~\ref{hov2:tab-210} summarizes the performance metrics ---
VMT, VHT and delay --- computed by simulation versus those values obtained from PeMS.
The PeMS data come from both SR-134 East and I-210 East, and VMT, VHT and delay
values are computed as sums of the corresponding values from these two
freeway sections.
Delay values are computed in vehicle-hours for those vehicles traveling
slower than 45 mph.

\begin{table}[ht]
\centering
\begin{tabular}{ l r r }
    \toprule
 Performance metric & Simulation result & PeMS data  \\ \midrule
GP Lane VMT & 2,017,322  & - \\ 
Managed Lane VMT & 378,485 & - \\ 
Total VMT & 2,395,807 & 414,941 + 2,006,457 = 2,421,398 \\ 
GP Lane VHT & 33,533 & - \\ 
Managed Lane VHT & 6,064 & - \\ 
Total VHT & 39,597 & 6,416 + 36,773 = 43,189 \\ 
GP Lane Delay (hr) & 3,078 & - \\ 
Managed Lane Delay (hr) & 584 & - \\ 
Total Delay (hr) & 3,662 & 1 + 3,802 = 3,803 \\ \bottomrule
\end{tabular}
\caption{Performance metrics for SR-134 East/ I-210 East.}
\label{hov2:tab-210}
\end{table}

\subsection{Discussion}
As we mention in section \ref{hov2:sec_calibration}, a common calibration goal for traffic modelers is to create a simulation that can accurately recreate a ``base case'' of a typical real-world day's traffic patterns; this calibrated ``base case'' can then be tweaked to predict the system-wide effects of counterfactual events such as a global demand increase, a localized lane closure, etc.

One traditional criterion for evaluating whether a freeway model captures this ``base case'' is whether the simulation predicts congestion at the same time and the same place as in real-world data (and, if the dynamic behaviors such as the rate of the buildup of the queue and the rate of queue discharge are similar).
In this congestion-locality criterion, our simulation case studies perform well.
For the case study of I-680 North, examining Figures \ref{hov2:fig-680-density} through \ref{hov2:fig-680-speed} we see that our models predict congestion originating at the bottleneck locations and propagating to the extents identified in the CSMP study \citep{system_metrics_group_inc._contra_2015}.
For the SR-134 East / I-210 East, we can compare the reference PeMS loop data speed contour (Figure \ref{hov2:fig-210-pems-speed}) to our simulations (Figures \ref{hov2:fig-210-density}-\ref{hov2:fig-210-speed}).
The PeMS data show and out simulations predict congestion in both the managed lane and the GP lanes that originates at a bottleneck at N. Azusa Avenue at roughly 15:00, propagates backward in space until reaching its maximum spatial extent roughly at postmile 29 at about 17:00, and then recedes until fully dissipating at roughly 19:00.
The managed lane data from the PeMS dataset (Figure \ref{hov2:fig-210-pems-speed}) has more identifiable wavefronts propagating backward and receding forward than the GP lanes, and those wavefront's speeds agree somewhat with those predicted by the simulation (unfortunately, a definitive value for the true wavefront speed is difficult to obtain due to the low resolution of the PeMS data).

Beyond the somewhat qualitative examination of whether the macroscopic congestion matters match reality, we can also evaluate quantitatively whether our simulation fits the available measurements from the site.
Figures \ref{hov2:fig-680-offramp-flow} and \ref{hov2:fig-210-offramp-flow} show how our model matched the offramp flow at the two identified bottlenecks of the case study sites.
Both figures show agreement between model and data.
Note that this ``offramp demand'' was an explicit calibration target (as described in section \ref{hov2:subsec-calibration-process}).
So, close matching of this is an explicit requisite for the calibrated model, and, it is necessary that the offramp flow here be satisfied, and the flow that does not take the offramp still produce the congestion patterns as discussed in the previous paragraph.
These two offramp flow figures can be interpreted as stating Neumann (flux) boundary conditions that our freeway PDE model must (and does) fulfill, in addition to the macro-scale congestion requirements.

Finally, Tables \ref{hov2:tab-680} and \ref{hov2:tab-210} show macroscopic freeway performance metrics predicted by our simulation and the comparable measured performance metrics.
For the three quantities for which measured performance metrics are available (total VMT, total VHT, and total delay), we can see that all percentage errors between our simulation and the measurements are at most roughly 10\%.
We consider this is a fairly good accuracy, given the inherent high noise of performance metrics computed from stationary sensors like the loop detectors used by PeMS \citep{jia_pems_2001,chen_detecting_2003}.
In particular, we find higher errors between the simulated performance metrics and the measured performance metrics for the SR-134 East / I-210 East case study, a location where single-loop detectors are present, than for the I-680 North case study, where mostly dual-loop detectors are present \citep{pems_california_2019} (single-loop detectors being intrinsically much more noisy than dual-loop detectors at measuring quantities like speed and flow \citep{jia_pems_2001}). 

\section{Conclusion}\label{hov2:sec_conclusion}
This paper presented the implementation of the modeling components for managed lane-freeway networks originally proposed in \citet{wright_macroscopic_2019}.
This included their integration into both a full macroscopic traffic model suitable for corridor-scale traffic simulation and analysis, as well as fitting in estimation methods for some of the harder-to-estimate parameters into the traditional iterative calibration approach for traffic models.

Our simulation results comparing our model and calibration results showed good agreement in two case studies, validating both our full-access and gated-access modeling techniques.
In the sequel to this paper series, we will further extend these results to include simulation and study of traffic control, with a reactive tolling controller on the managed lane.

\section*{Acknowledgements}\label{hov2:sec_acknowledgement}
We would like to express great appreciation to several of our colleagues.
To Elena Dorogush and Ajith Muralidharan for sharing ideas, and to
Gabriel Gomes and Pravin Varaiya for their
critical reading and their help in clarifying some theoretical issues.

This research was supported by the California Department of Transportation.
Earlier versions of some of this article's material previously appeared in the technical report \citet{horowitz_modeling_2016}.

\appendix
\section{A link Model}\label{hov:app_linkmodel}
\begin{figure}[ht]
\centering
\includegraphics[width=3in]{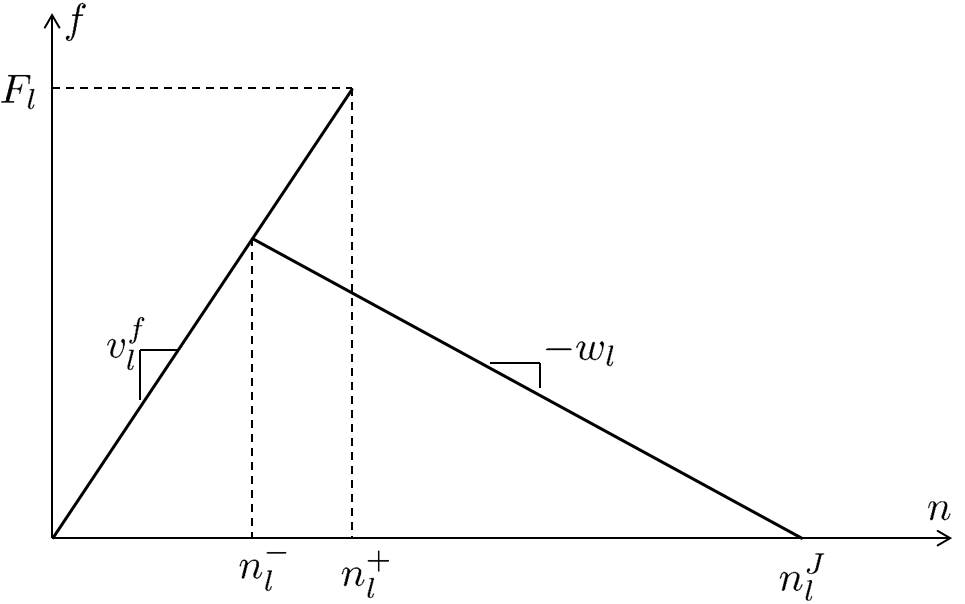}
\caption{The ``backwards lambda'' fundamental diagram.}
\label{hov:fig-fd}
\end{figure}
For the majority of this work, we remain agnostic as to the particular functional relationship between density $\rho_l$, demand $S_l$ (per commodity, $\rho_l^c$, $S_l^c$) and supply $R_l$, and flow $f_l$ (also called the fundamental diagram) used in our first-order macroscopic model.
For the simulation results presented in section \ref{hov2:sec_simulation}, we use a particular link model
shown in Figure~\ref{hov:fig-fd}.
This fundamental diagram captures the traffic hysteresis behavior with the ``backwards lambda'' shape often observed in detector data \citep{koshi_findings_1983}:
\begin{linenomath}
\begin{align}
S_l^c(t) =& v_l^f(t) \rho_l^c(t) \min \left\{1, \frac{F_l(t)}{v_l^f(t) \sum_{c=1}^C \rho_l^c(t)} \right\}, \;\;\;
S_l(t) = \sum_{c=1}^C S_l^c(t),
\label{hov:eq_lnctm_send_function} \\
R_l(t) =& \left( 1 - \theta_l(t) \right) F_l(t) + \theta_l(t) w_l(t)
  \left( \rho_l^J(t) - \sum_{c=1}^C \rho_l^c(t) \right), \label{hov:eq_lnctm_receive_function}
\end{align}
\end{linenomath}
where, for link $l$, $F_l$ is the capacity, $v_l^f$ is the free flow speed, $w_l$ is the congestion wave speed, $\rho_l^J$ is the jam density, and
$\rho_l^{-}=\frac{w_l\rho_l^J}{v_l^{f} + w_l}$ and $\rho_l^{+}=\frac{F_l}{v_l^{f}}$ are called the \emph{low} and \emph{high critical densities}, respectively.
As written here and used in this work, $F_l$, $v_l^f$, and $w_l$ are in units per simulation timestep.
The variable $\theta_l(t)$ is a congestion metastate of $l$, which encodes the hysteresis:
\begin{linenomath}
\begin{equation}
\theta_l(t) = 
\begin{cases}
0 & \rho_l(t)\leq \rho^-_l,\\
1 & \rho_l(t)> \rho^+_l,\\
\theta_l(t-1) & \rho_l^-<\rho_l(t)\leq \rho_l^+,
\end{cases}
\label{hov:eq_metastate}
\end{equation}
\end{linenomath}
where $\rho_l(t) = \sum_{c=1}^C \rho_l^c(t)$.

Examining~\eqref{hov:eq_metastate} and~\eqref{hov:eq_lnctm_receive_function}, we see that when a link's density goes above $\rho_l^{+}$ (i.e., when it becomes congested), its ability to receive flow is reduced until the density falls below $\rho_l^{-}$.

An image of~\eqref{hov:eq_lnctm_send_function}
and~\eqref{hov:eq_lnctm_receive_function} overlaid on each other,
giving a schematic image of the fundamental diagram,
is shown in Figure~\ref{hov:fig-fd}.
Unless $\rho_l^{-}=\rho_l^{+}$, when it assumes triangular shape,
the fundamental diagram is not a function of density alone (i.e., without $\theta_l(t)$):
$\rho_l(t)\in\left(\rho_l^{-}, \rho_l^{+}\right]$ admits two possible flow values.
 
\section{Dynamic Split Ratio Solver}\label{hov:app_splitratiosolver}
Throughout this article, we have made reference to a dynamic-system-based method for solving for partially- or fully-undefined split ratios from \citet{wright_node_2017}.
This split ratio solver is designed to implicitly solve the logit-based split ratio problem
\begin{equation}
  \beta_{ij}^c = \frac{\exp \left( \frac{\sum_{i=1}^M \sum_{c=1}^C S_{ij}^c}{R_j} \right)}{
  \sum_{j'=1}^N \exp \left( \frac{\sum_{i=1}^M \sum_{c=1}^C S_{ij'}^c}{R_{j'}} \right)}, \label{hov2:eq:splitratio_demandsupply}
\end{equation}
which cannot be solved explicitly, as the $S_{ij}^c$'s are also functions of the $\beta_{ij}^c$'s.
The problem \eqref{hov2:eq:splitratio_demandsupply} is chosen to be a node-local problem that does not rely on information from the link model (beyond supplies and demands), and is thus independent of the choice of link model \citep{wright_node_2017}.

The solution algorithm is as follows, reproduced from \citet{wright_node_2017}. More discussion is available in the reference.

\begin{itemize}
\item Define the set of commodity movements for which split ratios
are known as $\BB=\left\{\left\{i,j,c\right\}: \; \beta_{ij}^c \in [0,1]\right\}$,
and the set of commodity movements for which split ratios are to be
computed as $\OBB=\left\{\left\{i,j,c\right\}: \; \beta_{ij}^c \mbox{ are unknown}\right\}$.

\item For a given input link $i$ and commodity $c$ such that $S_i^c=0$,
assume that all split ratios are known: $\{i,j,c\}\in\BB$.\footnote{If split
ratios were undefined in this case, they could be assigned arbitrarily.}

\item Define the set of output links for which there exist unknown
split ratios as $V=\left\{j: \; \exists \left\{i,j,c\right\}\in\OBB\right\}$.

\item Assuming that for a given input link $i$ and commodity $c$,
the split ratios must sum up to 1, define the unassigned portion
of flow by $\obeta_i^c=1-\sum_{j:\{i,j,c\}\in\BB}\beta_{ij}^c$.

\item For a given input link $i$ and commodity $c$ such that there exists
at least one commodity movement $\{i,j,c\}\in\OBB$, assume $\obeta_i^c>0$, otherwise the
undefined split ratios can be trivially set to 0.

\item For every output link $j\in V$, define the set of input links
that have an unassigned demand portion directed toward this output link
by $U_j=\left\{i: \; \exists\left\{i,j,c\right\}\in\OBB\right\}$.

\item For a given input link $i$ and commodity $c$, define the set
of output links for which split ratios for which are to be computed as
$V_i^c = \left\{j: \; \exists i\in U_j\right\}$,
and assume that if nonempty, this set contains at least two elements,
otherwise a single split ratio can be trivially set equal to $\obeta_i^c$.

\item Assume that input link priorities are non-negative, $p_i\geq 0$,
	$i=1,\dots,M$, and $\sum_{i=1}^M p_i = 1$.

\item Define the set of input links with zero priority:
$U_{zp} = \left\{i:\;p_i=0\right\}$.
To enable split ratio assignment for inputs with zero priorities,
perform regularization:
\begin{equation}
\pt_i = p_i\left(1-\frac{|U_{zp}|}{M}\right)+ \frac{1}{M}\frac{|U_{zp}|}{M}=
p_i\frac{M-|U_{zp}|}{M} + \frac{|U_{zp}|}{M^2},
\label{hov2:priority_regularization}
\end{equation}
where $|U_{zp}|$ denotes the number of elements in set $U_{zp}$.
Expression~\eqref{hov2:priority_regularization} implies that the regularized
input priority $\pt_i$ consists of two parts:
(1) the original input priority $p_i$ normalized to the portion of 
input links with positive priorities; and
(2) uniform distribution among $M$ input links, $\frac{1}{M}$, 
normalized to the portion of input links with zero priorities.

Note that the regularized priorities $\pt_i> 0$, $i=1,\dots,M$, and $\sum_{i=1}^M \pt_i = 1$.
\end{itemize}

The algorithm for distributing $\obeta_i^c$ among the commodity movements
in $\OBB$ (that is, assigning values to the a priori unknown split ratios)
aims at maintaining output links as uniform in their demand-supply ratios as possible.
At each iteration $k$, two quantities are identified: $\mu^+(k)$, which is the largest \emph{oriented} demand-supply ratio produced by the split ratios that have been assigned so far, and $\mu^-(k)$, which is the smallest oriented demand-supply ratio whose input link, denoted $i^-$, still has some unclaimed split ratio.
Once these two quantities are found, the commodity $c^-$ in $i^-$ with the smallest unallocated demand has some of its demand directed to the $j$ corresponding to $\mu^-(k)$ to bring $\mu^-(k)$ up to $\mu^+(k)$ (or, if this is not possible due to insufficient demand, all such demand is directed).

To summarize, in each iteration $k$, the algorithm attempts to bring the smallest oriented demand-supply ratio $\mu^+(k)$ up to the largest oriented demand-supply ratio $\mu^-(k)$.
If it turns out that all such oriented demand-supply ratios become perfectly balanced, then the demand-supply ratios $(\sum_i \sum_c S_{ij}^c) / R_j$ are as well.

The algorithm is:

\begin{enumerate}
\item Initialize:
\begin{eqnarray*}
\tbeta_{ij}^c(0) & := & \left\{\begin{array}{ll}
\beta_{ij}^c, & \mbox{ if } \{i,j,c\}\in\BB,\\
0, & \mbox{ otherwise};\end{array}\right. \\
\obeta_i^c(0) & := & \obeta_i^c; \\
\Ut_j(0) & = &  U_j; \\
\Vt(0) & = & V; \\
k & := & 0,
\end{eqnarray*}
Here $\Ut_j(k)$ is the remaining set of input links with some unassigned demand,
which may be directed to output link $j$; and
$\Vt(k)$ is the remaining set of output links, to which the still-unassigned
demand may be directed.

\item If $\Vt(k)=\emptyset$, stop.
The sought-for split ratios are $\left\{\tbeta_{ij}^c(k)\right\}$,
$i=1,\dots,M$, $j=1,\dots,N$, $c=1,\dots,C$.

\item Calculate the remaining unallocated demand:
\[
\So_i^c(k) = \obeta_i^c(k) S_i^c, \;\;\; i=1,\dots,M, \;\; c=1,\dots,C.
\]

\item For all input-output link pairs,
calculate oriented demand:
\[ \St_{ij}^c(k) = \tbeta_{ij}^c(k) S_i^c. \]

\item For all input-output link pairs, calculate oriented priorities:
\begin{eqnarray}
\pt_{ij}(k) & = & \pt_i\frac{\sum_{c=1}^C \gamma_{ij}^c S_i^c}{
\sum_{c=1}^C S_i^c}
\label{hov2:eq_oriented_priorities_undefined_sr1} \\
\mbox{with} & & \nonumber \\
\gamma_{ij}^c(k) & = & \left\{\begin{array}{ll}
\beta_{ij}^c, & \mbox{ if split ratio is defined a priori: }
\{i,j,c\}\in\BB, \\
\tbeta_{ij}^c(k) + \frac{\obeta_i^c(k)}{|V_i^c|}, &
\mbox{ otherwise},\end{array}\right.
\label{hov2:eq_oriented_priorities_undefined_sr2}
\end{eqnarray}
where $|V_i^c|$ denotes the number of elements in the set $V_i^c$.
Examining the
expression~\eqref{hov2:eq_oriented_priorities_undefined_sr1}-\eqref{hov2:eq_oriented_priorities_undefined_sr2}, one can see that the
split ratios $\tbeta_{ij}^c(k)$, which are not fully defined yet,
are complemented with a fraction of $\obeta_i^c(k)$ inversely proportional
to the number of output links among which the flow of commodity $c$
from input link $i$ can be distributed.

Note that in this step we are using \emph{regularized} priorities $\pt_i$
as opposed to the original $p_i$, $i=1,\dots,M$.
This is done to ensure that inputs with $p_i=0$ are not ignored
in the split ratio assignment.

\item Find the largest oriented demand-supply ratio:
\[
\mu^+(k) = \max_{j} \max_{i}
\frac{\sum_{c=1}^C \St_{ij}^c(k)}{\pt_{ij}(k)R_j}\sum_{i\in U_j}\pt_{ij}(k).
\]

\item Define the set of all output links in $\Vt(k)$, where the minimum of
the oriented demand-supply ratio is achieved:
\[
Y(k) = \arg\min_{j\in\Vt(k)}\min_{i\in\Ut_j(k)}
\frac{\sum_{c=1}^C \St_{ij}^c(k)}{\pt_{ij}(k)R_j}
\sum_{i\in U_j}\pt_{ij}(k),
\]
and from this set pick the output link $j^-$ with the smallest
output demand-supply ratio (when there are multiple
minimizing output links, any of the minimizing output links
may be chosen as $j^-$):
\[
j^- = \arg\min_{j\in Y(k)}
\frac{\sum_{i=1}^M\sum_{c=1}^C\St_{ij}^c(k)}{R_j}.
\]

\item Define the set of all input links, where the minimum of
the oriented demand-supply ratio for the output link $j^-$ is achieved:
\[
W_{j^-}(k) = \arg\min_{i\in\Ut_{j^-}(k)}
\frac{\sum_{c=1}^C \St_{ij^-}^c(k)}{\pt_{ij^-}(k)R_{j^-}}
\sum_{i\in U_{j^-}}\pt_{ij^-}(k),
\]
and from this set pick the input link $i^-$ and commodity $c^-$
with the smallest remaining unallocated demand:
\[
\{i^-, c^-\} = \arg\min_{\begin{array}{c}
i\in W_{j^-}(k),\\
c:\obeta_{i^-}^c(k)>0\end{array}} \So_i^c(k).
\]

\item Define the smallest oriented demand-supply ratio:
\[
\mu^-(k) = 
\frac{\sum_{c=1}^C \St_{i^-j^-}^c(k)}{\pt_{i^-j^-}(k)R_{j^-}}
\sum_{i\in U_{j^-}}\pt_{ij-}(k).
\]
\begin{itemize}
\item If $\mu^-(k) = \mu^+(k)$, the oriented demands created by
the split ratios that have been assigned as of iteration $k$,
$\tbeta_{ij}^c(k)$, are perfectly balanced among the output links,
and to maintain this, all remaining unassigned split ratios should
be distributed proportionally to the allocated supply:
\begin{eqnarray}
\tbeta_{ij}^{c}(k+1) & = & \tbeta_{ij}^{c}(k) + 
\frac{R_j}{\sum_{j'\in V_{i}^{c}(k)}R_{j'}}
\obeta_{i}^{c}(k), \;\;\; 
c:~\obeta_i^c(k) > 0, \;\; i \in \Ut_j(k), \;\; j\in \Vt(k);
\label{hov2:eq_sr_assign_1}\\
\obeta_{i}^{c}(k+1) & = & 0, \;\;\; c:~\obeta_i^c(k) > 0,
\;\; i \in \Ut_j(k), \;\; j\in \Vt(k); \nonumber\\
\Ut_j(k+1) & = & \emptyset, \;\;\; j\in\Vt(k); \nonumber \\
\Vt(k+1) & = & \emptyset. \nonumber
\end{eqnarray}
If the algorithm ends up at this point, we have emptied $\Vt(k+1)$ and are done.
\item Else, assign:
\begin{eqnarray}
\Delta\tbeta_{i^-j^-}^{c^-}(k) & = & \min\left\{\obeta_{i^-}^{c^-}(k),\;\;
\left(\frac{\mu^+(k)\pt_{i^-j^-}(k)R_{j^-}}{
\So_{i^-}^{c^-}(k) \sum_{i\in U_{j^-}}\pt_{ij^-}(k)} -
\frac{\sum_{c=1}^C\St_{i^-j^-}^c(k)}{\So_{i^-}^{c^-}(k)}\right)\right\};
\label{hov2:eq_sr_assign_3}\\
\tbeta_{i^-j^-}^{c^-}(k+1) & = & \tbeta_{i^-j^-}^{c^-}(k) + 
\Delta\tbeta_{i^-j^-}^{c^-}(k);
\label{hov2:eq_sr_assign_4}\\
\obeta_{i^-}^{c^-}(k+1) & = & \obeta_{i^-}^{c^-}(k) -
\Delta\tbeta_{i^-j^-}^{c^-}(k); \label{hov2:eq_sr_assign_5} \\
\tbeta_{ij}^{c}(k+1) & = & \tbeta_{ij}^{c}(k) \mbox{ for }
\{i,j,c\}\neq\{i^-,j^-,c^-\}; \nonumber\\
\obeta_i^c(k+1) & = & \obeta_i^c(k) \mbox{ for } \{i,c\}\neq\{i^-,c^-\};
\nonumber \\
\Ut_j(k+1) & = & \Ut_j(k) \setminus \left\{i: \;
\obeta_{i}^c(k+1) = 0, \; c=1,\dots,C\right\}, \;\;\; j\in \Vt(k); \nonumber\\
\Vt(k+1) & = & \Vt(k) \setminus \left\{j: \; \Ut_j(k+1)=\emptyset\right\}.
\nonumber
\end{eqnarray}
In~\eqref{hov2:eq_sr_assign_3}, we take the minimum of the remaining unassigned
split ratio portion $\obeta_{i^-}^{c^-}(k)$ and the split ratio portion
needed to equalize $\mu^-(k)$ and $\mu^+(k)$.
To better understand the latter, the second term in $\min\{\cdot,\cdot\}$ can
be rewritten as:
\[
\frac{\mu^+(k)\pt_{i^-j^-}(k)R_{j^-}}{
\So_{i^-}^{c^-}(k) \sum_{i\in U_{j^-}}\pt_{ij^-}(k)} -
\frac{\sum_{c=1}^C\St_{i^-j^-}^c(k)}{\So_{i^-}^{c^-}(k)} =
\left(\frac{\mu^+(k)}{\mu^-(k)}-1\right)
\left(\sum_{c=1}^C\St_{i^-j^-}^c(k)\right)
\frac{1}{\So_{i^-}^{c^-}(k)}.
\]
The right hand side of the last equality can be interpreted as:
flow that must be assigned for input $i^-$, output $j^-$ and commodity $c^-$
to equalize $\mu^-(k)$ and $\mu^+(k)$ minus flow that is already assigned
for $\{i^-,j^-,c^-\}$, divided by the remaining unassigned portion
of demand of commodity $c^-$ coming from input link $i^-$.

In~\eqref{hov2:eq_sr_assign_4} and~\eqref{hov2:eq_sr_assign_5}, the assigned
split ratio portion is incremented and the unassigned split ratio portion is
decremented by the computed $\Delta\tbeta_{i^-j^-}^{c^-}(k)$.
\end{itemize}

\item Set $k := k+1$ and return to step 2.
\end{enumerate}

\bibliographystyle{abbrvnat}

\begin{thebibliography}{33}
\providecommand{\natexlab}[1]{#1}
\providecommand{\url}[1]{\texttt{#1}}
\expandafter\ifx\csname urlstyle\endcsname\relax
  \providecommand{\doi}[1]{doi: #1}\else
  \providecommand{\doi}{doi: \begingroup \urlstyle{rm}\Url}\fi

\bibitem[{Caltrans Office of Project Development
  Procedures}(2011)]{how_caltrans_builds_projects}
{Caltrans Office of Project Development Procedures}.
\newblock How {{Caltrans Builds Projects}}, Aug. 2011.
\newblock URL
  \url{http://www.dot.ca.gov/hq/tpp/index_files/How_Caltrans_Builds_Projects_HCBP_2011a-9-13-11.pdf}.

\bibitem[Cassidy et~al.(2010)Cassidy, Jang, and
  Daganzo]{cassidy_smoothing_2010}
M.~J. Cassidy, K.~Jang, and C.~F. Daganzo.
\newblock The smoothing effect of carpool lanes on freeway bottlenecks.
\newblock \emph{Transportation Research Part A: Policy and Practice},
  44\penalty0 (2):\penalty0 65--75, Feb. 2010.
\newblock ISSN 09658564.
\newblock \doi{10.1016/j.tra.2009.11.002}.

\bibitem[Chang et~al.(2008)Chang, Wiegmann, Smith, and
  Bilotto]{chang_review_2008}
M.~Chang, J.~Wiegmann, A.~Smith, and C.~Bilotto.
\newblock A {{Review}} of {{HOV Lane Performance}} and {{Policy Options}} in
  the {{United States}}.
\newblock Report FHWA-HOP-09-029, {Federal Highway Administration}, 2008.

\bibitem[Chen et~al.(2003)Chen, Kwon, Rice, Skabardonis, and
  Varaiya]{chen_detecting_2003}
C.~Chen, J.~Kwon, J.~Rice, A.~Skabardonis, and P.~Varaiya.
\newblock Detecting {{Errors}} and {{Imputing Missing Data}} for
  {{Single}}-{{Loop Surveillance Systems}}.
\newblock \emph{Transportation Research Record}, 1855\penalty0 (1):\penalty0
  160--167, Jan. 2003.
\newblock ISSN 0361-1981.
\newblock \doi{10.3141/1855-20}.

\bibitem[Daganzo(1994)]{daganzo_cell_1994}
C.~F. Daganzo.
\newblock The cell transmission model: {{A}} dynamic representation of highway
  traffic consistent with the hydrodynamic theory.
\newblock \emph{Transportation Research Part B: Methodological}, 28\penalty0
  (4):\penalty0 269--287, Aug. 1994.
\newblock \doi{10.1016/0191-2615(94)90002-7}.

\bibitem[Daganzo(1995)]{daganzo_cell_1995}
C.~F. Daganzo.
\newblock The cell transmission model, {{Part II}}: {{Network}} traffic.
\newblock \emph{Transportation Research Part B: Methodological}, 29\penalty0
  (2):\penalty0 79--93, Apr. 1995.
\newblock ISSN 01912615.
\newblock \doi{10.1016/0191-2615(94)00022-R}.

\bibitem[Dervisoglu et~al.(2009)Dervisoglu, Gomes, Kwon, Muralidharan, Varaiya,
  and Horowitz]{dervisoglu_automatic_2009}
G.~Dervisoglu, G.~Gomes, J.~Kwon, A.~Muralidharan, P.~Varaiya, and R.~Horowitz.
\newblock Automatic calibration of the fundamental diagram and empirical
  obsevations on capacity.
\newblock \emph{88th Annual Meeting of the Transportation Research Board,
  Washington, D.C., USA}, 2009.

\bibitem[Farhi et~al.(2013)Farhi, {Haj-Salem}, Khoshyaran, Lebacque, Salvarani,
  Schnetzler, and De~Vuyst]{farhi_logit_2013}
N.~Farhi, H.~{Haj-Salem}, M.~Khoshyaran, J.-P. Lebacque, F.~Salvarani,
  B.~Schnetzler, and F.~De~Vuyst.
\newblock The {{Logit}} lane assignment model: First results.
\newblock In \emph{Transportation {{Research Board}} 92nd {{Annual Meeting
  Compendium}} of {{Papers}}}, 2013.

\bibitem[Fitzpatrick et~al.(2017)Fitzpatrick, Avelar, and
  Lindheimer]{fitzpatrick_operating_2017}
K.~Fitzpatrick, R.~Avelar, and T.~E. Lindheimer.
\newblock Operating {{Speed}} on a {{Buffer}}-{{Separated Managed Lane}}.
\newblock \emph{Transportation Research Record: Journal of the Transportation
  Research Board}, 2616\penalty0 (1):\penalty0 1--9, Jan. 2017.
\newblock ISSN 0361-1981, 2169-4052.
\newblock \doi{10.3141/2616-01}.

\bibitem[Fransson and Sandin(2012)]{fransson_framework_2012}
M.~Fransson and M.~Sandin.
\newblock \emph{Framework for {{Calibration}} of a {{Traffic State Space
  Model}}}.
\newblock Masters {{Thesis}}, Linkoping University, {Norrkoping, Sweden}, Oct.
  2012.

\bibitem[Horowitz et~al.(2016)Horowitz, Kurzhanskiy, Siddiqui, and
  Wright]{horowitz_modeling_2016}
R.~Horowitz, A.~A. Kurzhanskiy, A.~Siddiqui, and M.~A. Wright.
\newblock Modeling and {{Control}} of {{HOT Lanes}}.
\newblock Technical report, {Partners for Advanced Transportation Technologies,
  University of California, Berkeley}, {Berkeley, CA}, 2016.
\newblock URL \url{https://escholarship.org/uc/item/9mx3903c}.

\bibitem[Jang and Cassidy(2012)]{jang_dual_2012}
K.~Jang and M.~J. Cassidy.
\newblock Dual influences on vehicle speed in special-use lanes and critique of
  {{US}} regulation.
\newblock \emph{Transportation Research Part A: Policy and Practice},
  46\penalty0 (7):\penalty0 1108--1123, Aug. 2012.
\newblock ISSN 09658564.
\newblock \doi{10.1016/j.tra.2012.01.008}.

\bibitem[Jang et~al.(2012)Jang, Oum, and Chan]{jang_traffic_2012}
K.~Jang, S.~Oum, and C.-Y. Chan.
\newblock Traffic {{Characteristics}} of {{High}}-{{Occupancy Vehicle
  Facilities}}: {{Comparison}} of {{Contiguous}} and {{Buffer}}-{{Separated
  Lanes}}.
\newblock \emph{Transportation Research Record: Journal of the Transportation
  Research Board}, 2278:\penalty0 180--193, Dec. 2012.
\newblock ISSN 0361-1981.
\newblock \doi{10.3141/2278-20}.

\bibitem[Jia et~al.(2001)Jia, Chen, Coifman, and Varaiya]{jia_pems_2001}
Z.~Jia, C.~Chen, B.~Coifman, and P.~Varaiya.
\newblock The {{PeMS}} algorithms for accurate, real-time estimates of
  g-factors and speeds from single-loop detectors.
\newblock In \emph{2001 {{IEEE Intelligent Transportation Systems
  Proceedings}}}, pages 536--541. {IEEE}, 2001.
\newblock ISBN 0-7803-7194-1.
\newblock \doi{10.1109/ITSC.2001.948715}.

\bibitem[Koshi et~al.(1983)Koshi, Iwasaki, and Ohkura]{koshi_findings_1983}
M.~Koshi, M.~Iwasaki, and I.~Ohkura.
\newblock Some findings and an overview on vehicular flow characteristics.
\newblock In \emph{Proceedings of the 8th {{International Symposium}} on
  {{Transportation}} and {{Traffic Flow Theory}}}, volume 198, pages 403--426,
  1983.

\bibitem[Kurzhanskiy and Varaiya(2015)]{kurzhanskiy_traffic_2015}
A.~A. Kurzhanskiy and P.~Varaiya.
\newblock Traffic management: {{An}} outlook.
\newblock \emph{Economics of Transportation}, 4\penalty0 (3):\penalty0
  135--146, Sept. 2015.
\newblock ISSN 22120122.
\newblock \doi{10.1016/j.ecotra.2015.03.002}.

\bibitem[Lighthill and
  Whitham(1955{\natexlab{a}})]{lighthill_kinematic_1955_pt1}
M.~J. Lighthill and G.~B. Whitham.
\newblock On kinematic waves {{I}}. {{Flood}} movement in long rivers.
\newblock \emph{Proceedings of the Royal Society of London. Series A.
  Mathematical and Physical Sciences}, 229\penalty0 (1178):\penalty0 281--316,
  May 1955{\natexlab{a}}.
\newblock \doi{10.1098/rspa.1955.0088}.

\bibitem[Lighthill and
  Whitham(1955{\natexlab{b}})]{lighthill_kinematic_1955_pt2}
M.~J. Lighthill and G.~B. Whitham.
\newblock On kinematic waves {{II}}. {{A}} theory of traffic flow on long
  crowded roads.
\newblock \emph{Proceedings of the Royal Society of London. Series A.
  Mathematical and Physical Sciences}, 229\penalty0 (1178):\penalty0 317--345,
  May 1955{\natexlab{b}}.
\newblock ISSN 2053-9169.
\newblock \doi{10.1098/rspa.1955.0089}.

\bibitem[Liu et~al.(2011)Liu, Schroeder, Thomson, Wang, Rouphail, and
  Yin]{liu_analysis_2011}
X.~Liu, B.~Schroeder, T.~Thomson, Y.~Wang, N.~Rouphail, and Y.~Yin.
\newblock Analysis of {{Operational Interactions Between Freeway Managed
  Lanes}} and {{Parallel}}, {{General Purpose Lanes}}.
\newblock \emph{Transportation Research Record: Journal of the Transportation
  Research Board}, 2262:\penalty0 62--73, Dec. 2011.
\newblock ISSN 0361-1981.
\newblock \doi{10.3141/2262-07}.

\bibitem[Lou et~al.(2011)Lou, Yin, and Laval]{lou_optimal_2011}
Y.~Lou, Y.~Yin, and J.~A. Laval.
\newblock Optimal dynamic pricing strategies for high-occupancy/toll lanes.
\newblock \emph{Transportation Research Part C: Emerging Technologies},
  19\penalty0 (1):\penalty0 64--74, Feb. 2011.
\newblock ISSN 0968090X.
\newblock \doi{10.1016/j.trc.2010.03.008}.

\bibitem[Menendez and Daganzo(2007)]{menendez_effects_2007}
M.~Menendez and C.~F. Daganzo.
\newblock Effects of {{HOV}} lanes on freeway bottlenecks.
\newblock \emph{Transportation Research Part B: Methodological}, 41\penalty0
  (8):\penalty0 809--822, Oct. 2007.
\newblock ISSN 01912615.
\newblock \doi{10.1016/j.trb.2007.03.001}.

\bibitem[{Metropolitan Transportation
  Commission}(2019)]{metropolitan_transportation_commission_bay_2019}
{Metropolitan Transportation Commission}.
\newblock Bay {{Area Express Lanes}}, 2019.
\newblock URL \url{http://bayareaexpresslanes.org}.

\bibitem[Ngoduy and Maher(2012)]{ngoduy_calibration_2012}
D.~Ngoduy and M.~Maher.
\newblock Calibration of second order traffic models using continuous cross
  entropy method.
\newblock \emph{Transportation Research Part C: Emerging Technologies},
  24:\penalty0 102--121, Oct. 2012.
\newblock ISSN 0968090X.
\newblock \doi{10.1016/j.trc.2012.02.007}.

\bibitem[Obenberger(Nov/Dec 2004)]{obenberger_managed_2004}
J.~Obenberger.
\newblock Managed {{Lanes}}: {{Combining Access Control}}, {{Vehicle
  Eligibility}}, and {{Pricing Strategies Can Help Mitigate Congestion}} and
  {{Improve Mobility}} on the {{Nation}}'s {{Busiest Roadways}}.
\newblock \emph{Public Roads}, 68\penalty0 (3):\penalty0 48--55, Nov/Dec 2004.
\newblock URL
  \url{http://www.fhwa.dot.gov/publications/publicroads/04nov/08.cfm}.
\newblock Publication Number FHWA-HRT-05-002.

\bibitem[{PeMS}(2019)]{pems_california_2019}
{PeMS}.
\newblock California {{Performance Measurement System}}, 2019.
\newblock URL \url{http://pems.dot.ca.gov}.

\bibitem[Poole and Kotsialos(2012)]{poole_metanet_2012}
A.~Poole and A.~Kotsialos.
\newblock {{METANET Model Validation}} using a {{Genetic Algorithm}}.
\newblock \emph{IFAC Proceedings Volumes}, 45\penalty0 (24):\penalty0 7--12,
  Sept. 2012.
\newblock ISSN 14746670.
\newblock \doi{10.3182/20120912-3-BG-2031.00002}.

\bibitem[Poole and Kotsialos(2016)]{poole_second_2016}
A.~Poole and A.~Kotsialos.
\newblock Second order macroscopic traffic flow model validation using
  automatic differentiation with resilient backpropagation and particle swarm
  optimisation algorithms.
\newblock \emph{Transportation Research Part C: Emerging Technologies},
  71:\penalty0 356--381, Oct. 2016.
\newblock ISSN 0968090X.
\newblock \doi{10.1016/j.trc.2016.07.008}.

\bibitem[Richards(1956)]{richards_shock_1956}
P.~F. Richards.
\newblock Shock waves on the highway.
\newblock \emph{Operations Research}, 4\penalty0 (1):\penalty0 42--51, 1956.
\newblock \doi{10.1287/opre.4.1.42}.

\bibitem[{System Metrics Group,
  Inc.}(2015)]{system_metrics_group_inc._contra_2015}
{System Metrics Group, Inc.}
\newblock Contra {{Costa County I}}-680 {{Corridor System Management Plan Final
  Report}}.
\newblock Technical report, {Caltrans District 4}, 2015.
\newblock URL
  \url{http://dot.ca.gov/hq/tpp/corridor-mobility/CSMPs/d4_CSMPs/D04_I680_CSMP_Final_Revised_Report_2015-05-29.pdf}.

\bibitem[Treiber and Kesting(2013)]{treiber_traffic_2013}
M.~Treiber and A.~Kesting.
\newblock \emph{Traffic {{Flow Dynamics}}}.
\newblock {Springer Berlin Heidelberg}, {Berlin, Heidelberg}, 2013.
\newblock ISBN 978-3-642-32459-8 978-3-642-32460-4.
\newblock \doi{10.1007/978-3-642-32460-4}.

\bibitem[Wright et~al.(2017)Wright, Gomes, Horowitz, and
  Kurzhanskiy]{wright_node_2017}
M.~A. Wright, G.~Gomes, R.~Horowitz, and A.~A. Kurzhanskiy.
\newblock On node models for high-dimensional road networks.
\newblock \emph{Transportation Research Part B: Methodological}, 105:\penalty0
  212--234, Nov. 2017.
\newblock ISSN 01912615.
\newblock \doi{10.1016/j.trb.2017.09.001}.

\bibitem[Wright et~al.(2018)Wright, Horowitz, and
  Kurzhanskiy]{wright_dynamic-system-based_2018}
M.~A. Wright, R.~Horowitz, and A.~A. Kurzhanskiy.
\newblock A {{Dynamic}}-{{System}}-{{Based Approach}} to {{Modeling Driver
  Movements Across General}}-{{Purpose}}/{{Managed Lane Interfaces}}.
\newblock In \emph{Proceedings of the {{ASME}} 2018 {{Dynamic Systems}} and
  {{Controls Conference}} ({{DSCC}})}, page V002T15A003. {ASME}, Sept. 2018.
\newblock ISBN 978-0-7918-5190-6.
\newblock \doi{10.1115/DSCC2018-9125}.

\bibitem[Wright et~al.(2019)Wright, Horowitz, and
  Kurzhanskiy]{wright_macroscopic_2019}
M.~A. Wright, R.~Horowitz, and A.~A. Kurzhanskiy.
\newblock Macroscopic {{Modeling}}, {{Calibration}}, and {{Simulation}} of
  {{Managed Lane}}-{{Freeway Networks}}, {{Part I}}: {{Topological}} and
  {{Phenomenological Modeling}}.
\newblock \emph{to be submitted to Transportation Research Part B:
  Methodological}, 2019.
\newblock URL \url{https://arxiv.org/abs/1609.09470}.

\end{thebibliography}

\end{document}